\documentclass[11pt]{article}

\usepackage{color}
\usepackage{epsfig}
\usepackage{amssymb}
\usepackage{amsmath}
\usepackage{bm}
\usepackage{subcaption}
\usepackage{url}
\usepackage{algorithm}
\usepackage{algorithmic}

\usepackage{booktabs}

\usepackage{color}
\usepackage{epsfig}
\usepackage{amssymb}
\usepackage{amsmath}
\usepackage{amsfonts}
\usepackage[normalem]{ulem}

\usepackage{mwe}
\usepackage{graphbox}

\usepackage{array}
\newtheorem{remark}{Remark}[section]
\title{Sub-aperture SAR Imaging with Uncertainty Quantification}
\author{Victor~Churchill\thanks{Department of Mathematics, The Ohio State University, Columbus,
OH, 43201 USA. E-mail: churchill.77@osu.edu} \quad Anne~Gelb\thanks{Department of Mathematics, Dartmouth College, Hanover, NH, 03755 USA. Email: annegelb@math.dartmouth.edu.  This work is partially supported by AFOSR \#F9550-22-1-0411, NSF-DMS \#1502640, NSF-DMS \#1912685, ONR \#N00014-20-1-2595.}}

\graphicspath{{./figs/}}

\begin{document}
\maketitle

\abstract{
In the problem of spotlight mode airborne synthetic aperture radar (SAR) image formation, it is well-known that data collected over a wide azimuthal angle violate the isotropic scattering property typically assumed. Many techniques have been proposed to account for this issue, including both full-aperture and sub-aperture methods based on filtering, regularized least squares, and Bayesian methods. A full-aperture method that uses a hierarchical Bayesian prior to incorporate appropriate speckle modeling and reduction was recently introduced to produce samples of the posterior density rather than a single image estimate. This uncertainty quantification information is more robust as it can generate a variety of statistics for the scene. As proposed, the method was not well-suited for large problems, however, as the sampling was inefficient.  Moreover,  the method was not explicitly designed to mitigate the effects of the faulty isotropic scattering assumption. In this work we therefore propose a new sub-aperture SAR imaging method that uses a sparse Bayesian learning-type algorithm to more efficiently produce approximate posterior densities for each sub-aperture window.  These estimates may be useful in and of themselves,  or when of interest,  the statistics from these distributions can be combined to form a composite image. Furthermore, unlike the often-employed $\ell_p$-regularized least squares methods, no user-defined parameters are required.  Application-specific adjustments are made to reduce the typically burdensome runtime and storage requirements so that appropriately large images can be generated. Finally, this paper focuses on incorporating these techniques into SAR image formation process. That is, for the problem starting with SAR phase history data, so that no additional processing errors are incurred. The advantage over existing SAR image formation methods are clearly presented with numerical experiments using real-world data.
}

\section{Introduction}
\label{sec:introduction}
%\AG{I re-ordered the sentences a bit so read through to see they still make sense.}

Spotlight mode airborne synthetic aperture radar (SAR)\footnote{To be concise, we will refer to spotlight mode airborne SAR as just SAR from now on.} is a widely-used technology for surveillance and mapping due to its all-weather day-or-night imaging capabilities that make it favorable over optical imaging modalities in remote sensing, \cite{andersson2012fast,cheney2009,jakowatz2012spotlight}.  The data are typically collected in a circular flight pattern around the scene of interest and then processed to produce images that are used for purposes such as classification and identification.  It is therefore critical for practitioners to be able to rely on SAR images that clearly localize targets. 

\subsection*{Challenges in SAR imaging}
There are several well-known challenges in SAR imaging, which we discuss below, along with the existing state of the field on these issues. 

\begin{itemize}
\item \textbf{Sub-aperture imaging:} While it is straightforward to directly process the circularly-collected $360^o$-azimuth SAR phase history data into a single image estimate, i.e. so-called full-azimuth or full-aperture imaging, it forces the often used yet incorrect assumption that the imaging scene contains only isotropic scatterers. There is angular dependence of the scattered response, however. In particular, since the energy is scattered off objects in the front of the scene, then it does not mix evenly with the full scene. So called sub-aperture techniques, where the full azimuth data are grouped into smaller overlapping sub-aperture windows and then processed separately for image recovey  are often employed to alleviate this problem.  Each image recovery process typically employs filtering, regularized least squares, $\ell_1$ regularization, or Bayesian methods, followed by some mechanism (e.g. weighted averages) to aggregate the results from each sub-aperture recovery to form a final image.  A good review of sub-aperture techniques  may be found in \cite{ash2014wide}, with specific examples in \cite{Cetin2005,ccetin2014sparsity,moses2004wide,paulson,potter2010sparsity,sanders2017composite,stojanovic2008joint}. 
%Although other data collection schemes can be considered, \cite{moses2004wide}, this paper focuses on a single circular data collection, followed by processing of this full-azimuth data into overlapping or non-overlapping segments.
%A final composite image for the scene can also be recovered by taking the maximum, a (weighted) average, or some other appropriate combination of the sub-aperture images.
%In \cite{moses2004wide}, the idea of composite imaging for SAR from multiple sub-apertures is formed using a simple pseudoinverse reconstruction for each sub-aperture image. In \cite{cetin2005sar}, sparsity-encouraging $\ell_1$-regularized reconstruction, \cite{potter2010sparsity}, is used to enhance the results. Finally, \cite{ash2014wide} provides a review of the literature, distinguishing between sub-aperture and full aperture approaches. Indeed in \cite{sanders2017composite} a more sophisticated approach that exploited commonalities between neighboring aperture windows was shown to be very effective for improving image fidelity and reducing speckle. In this paper we develop a composite image formation approach based on the sampling method introduced in \cite{CGspeckle}, for which the model directly incorporates speckle for each aperture window. In so doing, we demonstrate a more accurate SAR image recovery process than current composite formulations that can also provide valuable uncertainty quantification information.

\item \textbf{Speckle:} SAR is a coherent imaging system, meaning that both the collected data and the reflectivity image are complex-valued. While typically only the magnitude is viewed, the phase information should not be neglected in the image formation process, and is important for downstream tasks such as target position and amplitude estimation, \cite{moore2017using,moore2018characterization}. All coherent imaging modalities are affected by speckle, a multiplicative-noise-like phenomenon which, although is in fact signal, causes grainy-looking images and hence it is desirable to remove it. Existing methods for SAR image reconstruction from phase history data usually do not directly address speckle and {post-processing operations like smoothing and filtering are typically necessary, see e.g. \cite{argenti2013tutorial,chierchia2017multitemporal,cozzolino2013fast,daoui2020stable,lee1994speckle,parrilli2011nonlocal}}. Basic, fast methods that rely on an inverse non-uniform fast Fourier transform (NUFFT), \cite{gorham2010sar}, provide no speckle reduction, while sparsity-based methods that rely on $\ell_p$ regularization in some domain\footnote{Speckle reduction is often also addressed using denoising techniques such as total variation (TV) regularization, either used directly of through a joint sparsity scheme \cite{potter2010sparsity,rudin1992nonlinear,sanders2017composite}.} \cite{archibald2016image,ccetin2001feature,cetin2005sar,ccetin2014sparsity,scarnati2018joint,scarnati2018recent,potter2010sparsity,samadi2011sparse}, disregard the physical meaning of speckle and instead choose to place penalties on approximate pixel magnitude values. Conflating speckle with the usual additive noise in this way makes parameter selection for the $\ell_1$ regularization penalty term very difficult (and essentially without physical meaning) in practice, although some methods have tried to tackle estimating this parameter \cite{batu2011parameter}. Recently, \cite{CGspeckle} incorporated the fully-developed speckle model as an image prior within a hierarchical Bayesian framework, and in doing so properly characterized the speckle as part of the signal, which could then be appropriately reduced, for example using a sparse Bayesian learning (SBL) technique \cite{tipping2001sparse}. The numerical tests performed in \cite{CGspeckle}  confirmed that using this model yielded improved speckle reduction over commonly used compressive sensing techniques while also providing uncertainty quantification (see below). However, due to long run-time, this method did not consider sub-aperture imaging and the correspondingly required processing of more than one aperture window.

\item \textbf{Uncertainty quantification:} For the most part the existing sub-aperture and full azimuth SAR image formation methods produce  a single image product, typically a maximum likelihood or maximum \emph{a posteriori} point estimate, that approximates the unknown ground truth. Predictions from these methods are not probabilistic, and therefore provide no information about the statistical confidence with which we can trust the features in the resulting images, e.g., which are more likely actual objects in the scene and which are more likely attributed to speckle or noise. This makes forming reliable images difficult, particularly in SAR where even many synthetically-created examples have unknown true reflectivity.\footnote{We point out that there are some instances for which benchmarks for despeckling have been established, \cite{di2013benchmarking}. However, these tests operate on simulated magnitude-only images that have already been formed, while the focus of this paper is on reconstructing complex-valued images from real-world phase history data.} In \cite{CGspeckle}, an MCMC-based method was used to sample the posterior density of a hierarchical Bayesian model for full-azimuth SAR image reconstruction from phase history data, giving much more robust information about the distribution of the image to affirm the confidence with which each estimate can be trusted. Sampling-based methods using this same prior structure have been developed to quantify uncertainty in basic real-valued linear inverse problems such as image reconstruction, see e.g. \cite{bardsley2012mcmc}, and have also been applied to SAR imaging tasks such as moving target inference, \cite{newstadt2014moving}, passive SAR image reconstruction, \cite{wu2015high}, and speckle noise model selection, \cite{karakucs2018generalized}. In addition, several SAR image reconstruction methods \cite{duan2015pattern,wu2011compressive,xu2012bayesian,xue2009sar} have leveraged SBL estimation procedures, \cite{tipping2001sparse}, or Bayesian compressed sensing \cite{ji2008bayesian} to analyze the same posterior.

\end{itemize}

The purpose of this paper therefore is to address these three issues by introducing an efficient sub-aperture SAR imaging method that models and reduces speckle and returns an approximate posterior density, allowing for uncertainty quantification. Significantly, we are able to address these tasks within the process of reconstructing SAR images directly from phase history data, as opposed to relying on altering images that have already been reconstructed or otherwise processed. 

%\AG{It looks like something is missing here.}  is used to approximate the sub-aperture densities.

\subsection*{Contributions}%\label{subsec:contributions}

This investigation provides several significant contributions to the SAR image formation literature. First, to the best of our knowledge, sub-aperture approaches for wide-angle SAR imaging have previously been limited to point estimates. To this end, our proposed method  provides approximate posterior densities for each sub-aperture image based on a deterministic estimation procedure \cite{glaubitz2022generalized} to analyze the constructed hierarchical Bayesian prior. Second, while many sub-aperture methods have been shown to yield highly resolved SAR images, \cite{ash2014wide,Cetin2005,ccetin2014sparsity,moses2004wide,paulson,potter2010sparsity,sanders2017composite,stojanovic2008joint},  whether through a direct prior on each sub-aperture image or a joint sparsity approach, these techniques all require some user selected  thresholds and/or parameters, which makes them less robust to modifications within the data. By contrast, the proposed method is automated, meaning that it does not require any parameter tuning for different data sets. While some methods have been proposed to aid in selection of the regularization parameters, e.g.~ \cite{batu2011parameter,VBJS_Speckle}, the task remains challenging in practice and at the very least requires some extra processing. The method we propose is also able to incorporate TV regularization (as well as higher order TV, and most other regularization operators, e.g. wavelets) into its hierarchical Bayesian prior. The ability to regularize in other domains exhibits the general flexibility of the Bayesian model, whereby additional priors can be added to account for a variety of constraints, such as phase error correction \cite{sanders2017combination,wu2015sparse,scarnati2018joint,su2016joint,onhon2011sparsity,xu2011bayesian}. To the best of our knowledge, this is the first time an SBL-type prior has been used with TV for SAR image formation. Third, we demonstrate that this novel approach is also significantly faster than other sparsity ($\ell_1$-regularized least squares) methods. Furthermore, among other subjective advantages in the actual appearance of the image, our technique increases image contrast and reduces speckle. Finally, a challenging real-world data example is considered to demonstrate the new methodology, and supports its use. Real-world SAR image formation is a very large problem, and we make application-specific efficient storage and runtime adjustments due to the prohibition of traditional matrix-based methods for linear inverse problems as even storing dense matrices of the necessary size is problematic. Parallel processing of the sub-aperture images is also included in the implementation.

\subsection*{Organization}
The rest of this paper is organized as follows.  Section \ref{sec:background} gives necessary background on SAR imaging and existing estimation techniques. Section \ref{sec:compositeSAR} derives the hierarchical Bayesian prior of \cite{tipping2001sparse} from scratch, emphasizing the incorporation of coherent imaging, speckle, and sparsity using conjugate priors, and then explains the new algorithm and its properties.  Section \ref{sec:results} shows a real-world example using the Air Force Research Laboratory's GOTCHA Volumetric Data Set 1.0, \cite{casteel2007challenge}. Some concluding remarks and ideas for future directions are provided in Section \ref{sec:conclusion}.

\section{Background}\label{sec:background}

\begin{figure}[h]
\centering
\includegraphics[width=.30\textwidth]{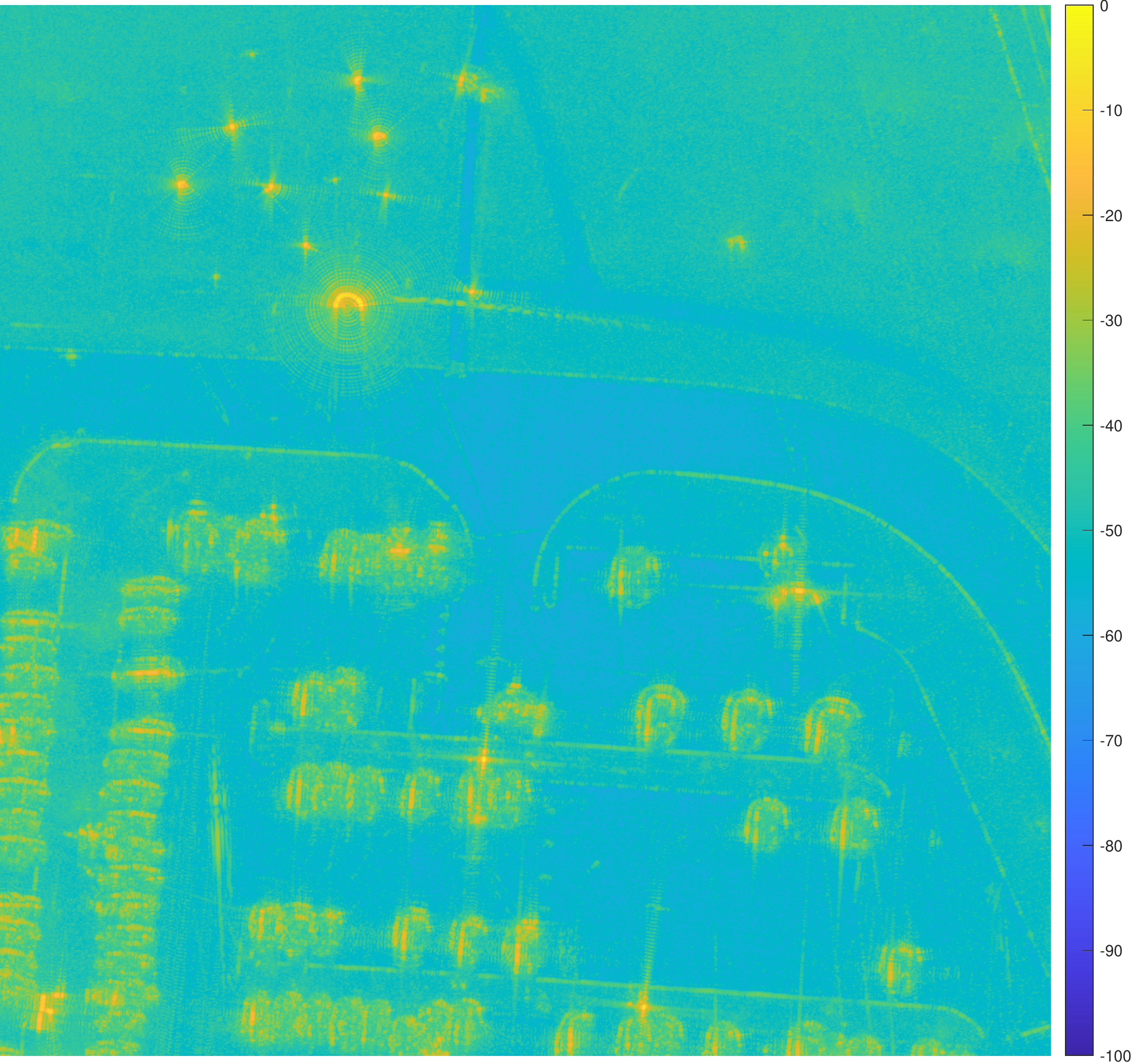}
\includegraphics[width=.30\textwidth]{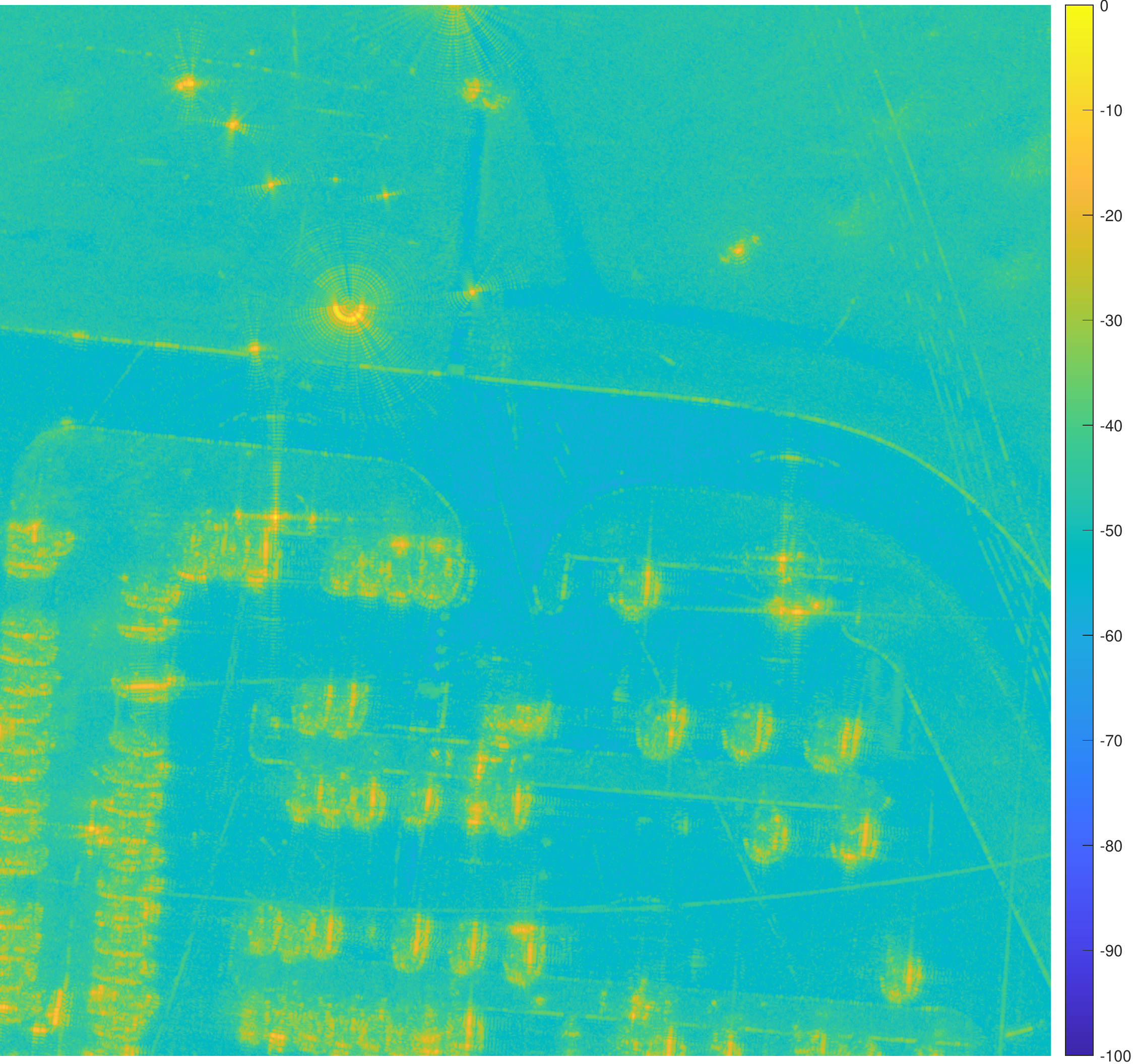}
\caption{Parking lot images formed using opposite halves of full azimuth coverage ($2048\times2048$ pixels, NUFFT image formation).}
\label{fig:GOTCHA_180}
\end{figure}

\begin{figure}[h]
\centering
\includegraphics[width=.30\textwidth]{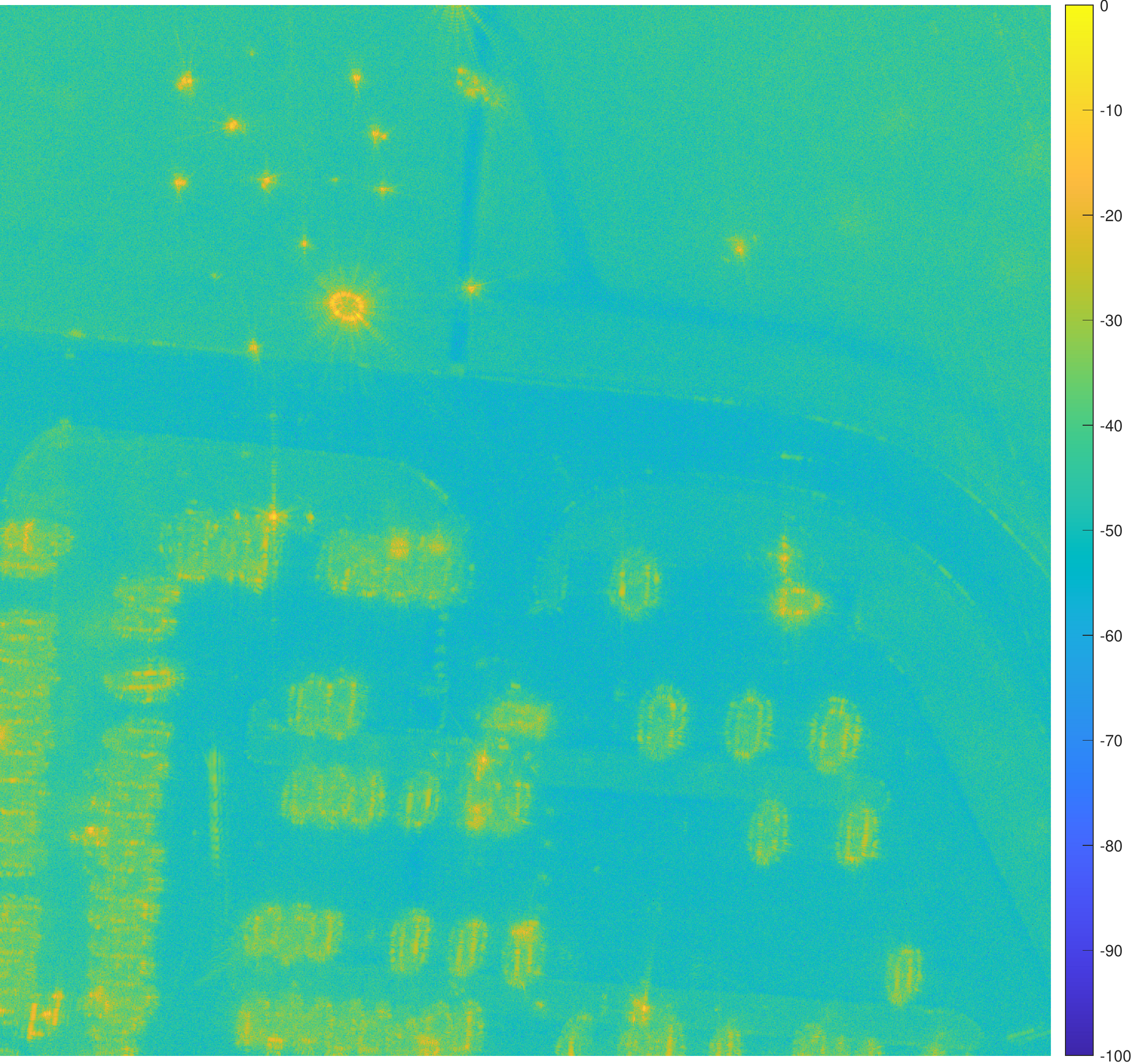}
\includegraphics[width=.30\textwidth]{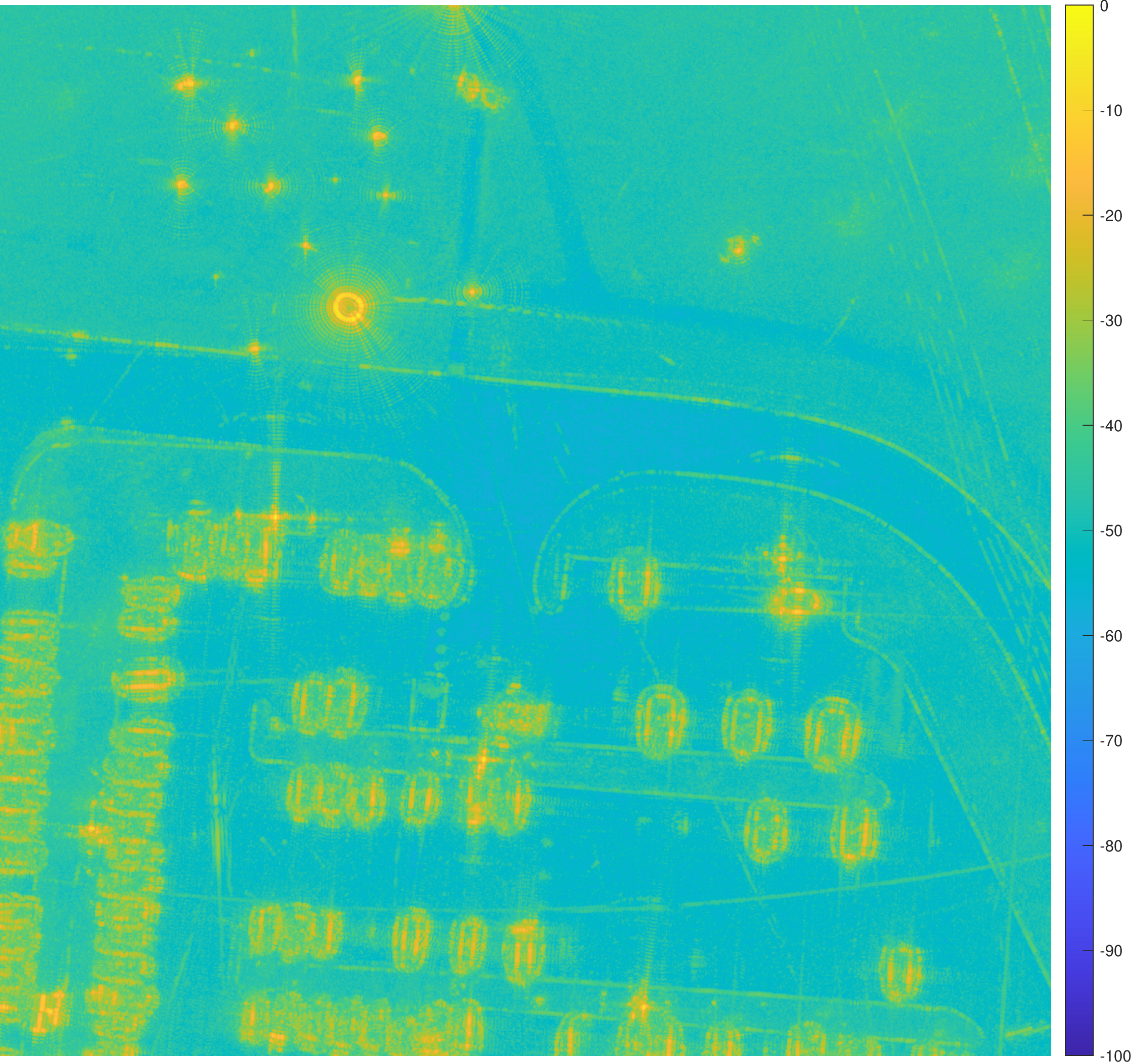}
\caption{Parking lot images formed via (left) full-aperture and (right) sub-aperture methods ($2048\times2048$ pixels, NUFFT image formation with 60 sub-apertures each spanning $8^o$ with $2^o$ of overlap).}
\label{fig:GOTCHA_widecomp}
\end{figure}

As noted in Section \ref{sec:introduction}, although it is straightforward to directly process the circularly-collected $360^o$-azimuth SAR phase history data into a single full-azimuth image estimate, it forces the often used yet incorrect assumption that the imaging scene contains only isotropic scatterers.  By using opposite halves of the full azimuth to recover the same underlying scene, Figure \ref{fig:GOTCHA_180} demonstrates that this assumption is clearly false.  It is reasonable to assume, however, that the information received is the same within a small aperture window.   Therefore the {\em sub-aperture imaging approach}, \cite{cetin2005sar,moses2004wide,sanders2017composite}, which forms a SAR image by combining a finite number of individual reconstructions coming from small overlapping aperture windows of data, helps to mitigate the effects of conflicting information resulting from the anisotropic nature of the scatterers.   Observe in Figure \ref{fig:GOTCHA_widecomp}(right) that simply  applying a non-uniform Fourier transform (NUFFT) to each sub-aperture window and then combining the results by taking the maximum modulus over the sub-apertures to form a single image clearly yields more localized features.  

\subsection{SAR modeling}
\label{sec:SARmodel}
The ensuing description of the SAR image formation inverse problem below generally follows the development in \cite{jakowatz2012spotlight,scarnati2018joint}. Let $f:\Omega \rightarrow \mathbb{C}$ denote the two-dimensional reflective scene of scattering objects that we want to recover, where $f$ is defined over
$$
\Omega := \{ (x,y)\in\mathbb{R}^2 | x^2 + y^2 \leq U^2\}.
$$
At a particular position in the sensing process, indicated by an azimuth angle $\theta\in[0,2\pi]$, the transmitted linear FM chirp mixes with the scene in a way that depends upon $\theta$, the angle from which the chirp is emitted.\footnote{In practice there is also a relevant angle of elevation which is not critical to the development of our method.}   In the far field case, once the transmitted signal reaches the scene it has essentially a planar wave front, and thus the points in the scene along each line perpendicular to the direction of the chirp all mix with the same values.  Hence the two-dimensional setup is often simplified to a one-dimensional process by compressing the scatterers along each of these lines to a single point.  This compression is commonly referred to as the projection or Radon transform of $f$ at the angle $\theta$, and is denoted $p:\Omega\rightarrow\mathbb{R}^2$. It can be expressed mathematically as
\begin{equation}\label{eq:radon}
p(\theta,u) = \iint\displaylimits_{x^2+y^2\le U^2} f(x,y) \delta(u - x\cos\theta-y\sin\theta ) \, dx \, dy,
\end{equation}
{where $\delta$ is the Dirac delta function and $u \in \mathbb{R}$ is the slant range position.}

The linear FM chirp that is transmitted and mixed with the scene is the real part of
\begin{equation}\label{eq:chirp}
 s(t) = \begin{cases}
         e^{i(\omega t + \alpha t^2)} , & |t|\le \frac{T}{2}\\
         0, & \text{otherwise}
        \end{cases},
\end{equation}
where $\omega$ is the carrier frequency, $2\alpha$ is the chirp rate, and $T>0$ is the pulse duration.  This chirp signal mixes with the scene to yield reflected signals of the form
\begin{equation}\label{eq:chirp-conv}
r(\theta,t) =   \int_{-U}^{U}\text{Re}\left\{ p(\theta,u) s(t-\tau_0 - \tau(u)) \right\} \, du ,
\end{equation}
where $\tau_0$ is the round trip time required for the chirp to travel to the scene center and  $\tau(u)$  is the additional travel time for any particular position in the scene $u$.  If $R$ is the distance from the transmitter/receiver to the scene center and $c$ is the speed of light in a vacuum, we have $\tau_0 = 2R/c$   and $\tau(u) = 2u/c$.

A deramping process is implemented to  extract approximate instantaneous frequency information (i.e.~the classical Fourier transform of $f$) from the chirp response, ultimately yielding the approximation\footnote{More details may be found in \cite{jakowatz2012spotlight,scarnati2018joint}.}
\begin{equation}\label{eq:fourier}
\hat{f}_{\theta}(t) :=  \int_{-U}^{U} p(\theta,u) e^{-iku} \, du  \approx r(\theta,t),
\end{equation}
where the spatial frequencies $k$, measured in cycles/meter, are given by
\begin{equation}
\label{eq:spat_freq_AF}
k =  k(t) := \frac{2}{c}(\omega + 2\alpha(t-\tau_0)).
\end{equation}
This approximation makes several assumptions about the transmitted signal, the scene, and how accurately the data are measured.   For example, as already noted, in far field spotlight SAR, the distance $R$ from the transmitter/receiver to the scene center is far enough so that the transmitted signal has essentially a planar wave front (i.e.~any curvature of the scene geometry can be safely ignored). Moreover, the chirp rate $\alpha$ is small enough so that higher order terms can be ignored in the approximation of \eqref{eq:fourier}. A third assumption often made is that the round trip propagation time is exactly known.  This is generally not the case, and this lack of precision causes a phase error in the signal recovery.  Autofocusing is commonly employed to reduce the shearing effect caused by the phase error. We do not discuss the ramifications of phase error in the current investigation, but point readers to \cite{ash2012autofocus,scarnati2018joint,ugur2012sar,sanders2017combination,wu2015sparse,scarnati2018joint,su2016joint,onhon2011sparsity,xu2011bayesian} for general information on autofocusing techniques.  Finally, \eqref{eq:spat_freq_AF} assumes that the scene scatterers in spotlight SAR are isotropic, meaning that the reflection is the same regardless of the azimuth angle. Figures \ref{fig:GOTCHA_180} and \ref{fig:GOTCHA_widecomp} respectively demonstrate the inaccuracy in the image recovery due to this faulty assumption and how using a sub-aperture  approach helps to alleviate this issue.

The projection slice theorem \cite{jakowatz2012spotlight} is used to conveniently rewrite  \eqref{eq:fourier} as
\begin{equation}
\label{eq:projectionslice} 
\hat{f}_{\theta}(t) =  \int_{-U}^{U} p(\theta,u) e^{-iku} \, du = \iint\displaylimits_{x^2+y^2\le U^2} f(x,y) e^{-ik(x\cos\theta+y\sin\theta)}dxdy.
\end{equation}
Hence forming an image from SAR phase history data can be modeled as the inverse of a continuous non-uniform Fourier transform.  

To discretize the problem, let temporal frequency values be given by $t_m$ for $m = 1,...,M$, and a set of azimuth angles by $\theta_l$ for each sub-aperture $l = 1,...,L$. From \eqref{eq:spat_freq_AF} we can define
\begin{equation}
\label{eq:spat_freq_disc}
k_m = \frac{2}{c}(\omega + 2\alpha(t_m-\tau_0)), \hspace{.2in}m = 1,...,M,
\end{equation}
as the discretized spatial frequencies, leading to the linear system model for each sub-aperture image formation
\begin{align}\label{eq:discretemodel}
\mathbf{\hat{f}}^{(l)} = \mathbf{F}^{(l)}\mathbf{f}^{(l)} + \mathbf{n}^{(l)}, \quad l=1,\ldots,L.
\end{align}
The discrete forward operator  $\mathbf{F}^{(l)}: \mathbb{C}^{N} \rightarrow \mathbb{C}^{M}$, $l=1,\ldots,L$, is the two-dimensional discrete non-uniform Fourier transform matrix for the sub-aperture data collection that maps modeled by \eqref{eq:projectionslice} that maps the concatenated $l$-th sub-aperture reflectivity image $\mathbf{f}^{(l)}\in \mathbb{C}^{N}$ to the corresponding vertically-concatenated phase history data $\mathbf{\hat{f}}^{(l)}\in\mathbb{C}^{M}$, where $N$ is the number of pixels in the each image, $M$ is the length of the data in each window, and $L$ is the number of sub-aperture windows.\footnote{In fact it is possible to have $M = M(l)$, $l = 1,\ldots,L$, but for ease of presentation we choose the number of frequencies in each window to remain constant.}  Note that \eqref{eq:discretemodel} is equivalent to the full-aperture case when $L = 1$, i.e. a single aperture window with full azimuth information.  

Since it is reasonable to assume that within a small aperture the scatterers in the scene are isotropic, the method proposed in this investigation uses the sub-aperture imaging approach of dividing the full azimuth data into sub-apertures which may or may not overlap.
Moreover, based on the speckle model, \cite{jakowatz2012spotlight}, we can also assume in \eqref{eq:discretemodel} that $\mathbf{n}^{(l)}$ is complex noise that is circularly-symmetric white Gaussian distributed. That is, $\mathbf{n}^{(l)}_i\sim\mathcal{CN}(0,1/\beta^{(l)})$ i.i.d. for all pixels $i$, where $(\beta^{(l)})^{-1}>0$ is the noise variance. This yields the Gaussian likelihood function
\begin{align}\label{eq:likelihood}
p(\mathbf{\hat{f}}^{(l)}| \mathbf{f}^{(l)}, \beta^{(l)}) \propto (\beta^{(l)})^{M} \exp\left(-\frac{\beta^{(l)}}{2}\|\mathbf{\hat{f}}^{(l)}-\mathbf{F}^{(l)}\mathbf{f}^{(l)}\|_2^2\right),\quad\quad l = 1,\ldots, L,
\end{align}
which measures the goodness of fit of the model \eqref{eq:discretemodel}, where $||\mathbf{g}||^2 = \mathbf{g}^\ast\mathbf{g}$ with $\mathbf{g}^\ast$ the conjugate transpose of $\mathbf{g}$. The initial objective is to infer $\mathbf{f}^{(l)}$ from $\mathbf{\hat{f}}^{(l)}$ in each sub-aperture, and then combine each of the $L$ recovered windowed images into a single image (if desired).   Note that by using the discrete Fourier transform in \eqref{eq:discretemodel} we introduce both aliasing error and the Gibbs phenomenon. Given sufficient resolution, the magnitudes of these errors are within the range of additive noise so we do not consider them further.

\subsection{Estimation techniques}
\label{sec:estimate}
One straightforward way to estimate each $\mathbf{f}^{(l)}$ from $\mathbf{\hat{f}}^{(l)}$ is to maximize the likelihood function. From the Gaussian likelihood in \eqref{eq:likelihood}, this estimate is
\begin{align}\label{eq:MLestimate}
\begin{split}
\mathbf{f}^{(l)}_{ML} &= \arg\max_\mathbf{f} \left\{ p(\mathbf{\hat{f}}^{(l)}|\mathbf{f},\beta)\right\} \\
&= \arg\min_\mathbf{f} \left\{ \|\mathbf{\hat{f}}^{(l)} - \mathbf{F}^{(l)}\mathbf{f}\|_2^2\right\}, \quad\quad l = 1,\ldots, L.
\end{split}
\end{align}
Assuming approximate orthogonality of $\mathbf{F}^{(l)}$, we have $\mathbf{f}^{(l)}_{ML}\approx(\mathbf{F}^{(l)})^\ast\mathbf{\hat{f}}^{(l)}$, which only requires an inverse non-uniform fast Fourier transform (NUFFT) application to invert the data, \cite{fessler2003nonuniform,lee2005type,greengard2004accelerating}. Specifically, an individual reflectivity image can be found by interpolating the typically polar grid of measured samples in frequency space to an equally spaced rectangular grid, then computing an inverse \emph{uniform} fast Fourier transform, \cite{andersson2012fast}.  While the NUFFT has the advantage of being computationally efficient, the noisy data and model error can still degrade image quality. Hence an improvement is generally needed. Each image in Figures \ref{fig:GOTCHA_180} and \ref{fig:GOTCHA_widecomp} was formed using the above NUFFT method. We highlight the differences between Figure \ref{fig:GOTCHA_widecomp}(left), which uses $L=1$ full-azimuth window,  and Figure \ref{fig:GOTCHA_widecomp}(right),  which shows the maximum modulus composite image of $L=60$ sub-aperture windows of $8^o$ with $2^o$ of overlap.

\begin{figure}[h]
\centering
\includegraphics[width=.3\textwidth]{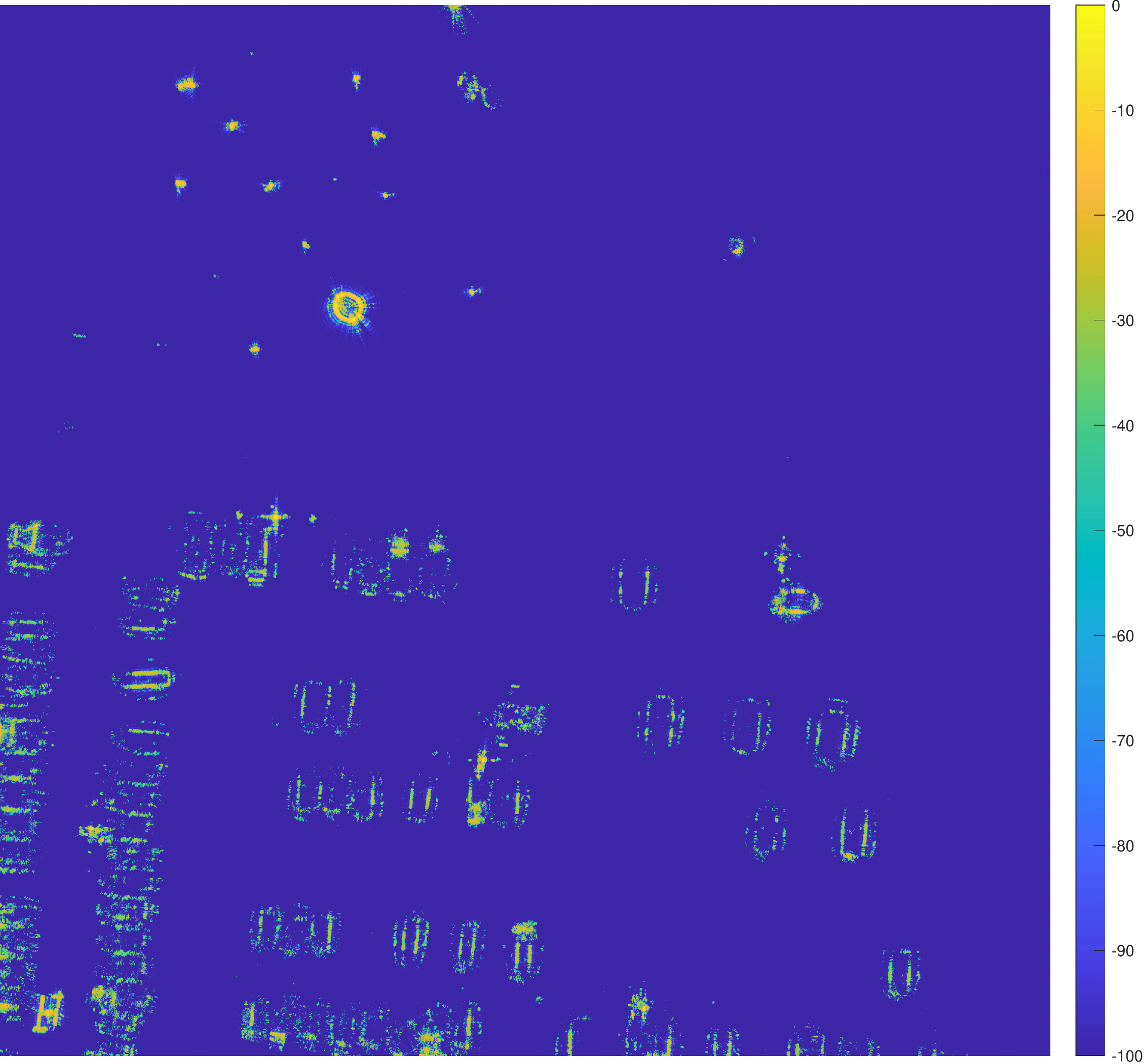}
\includegraphics[width=.3\textwidth]{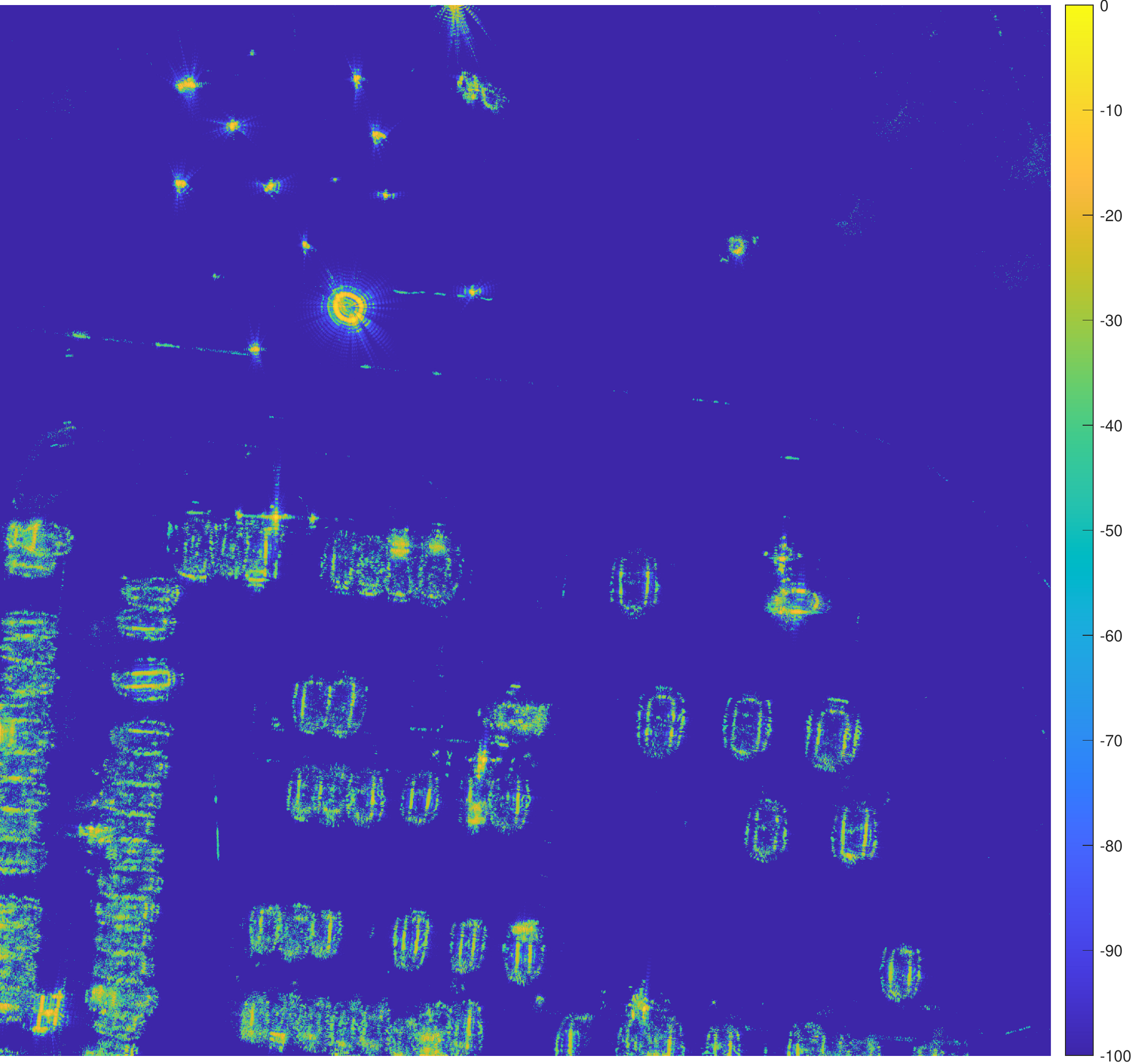}\\
\includegraphics[width=.3\textwidth]{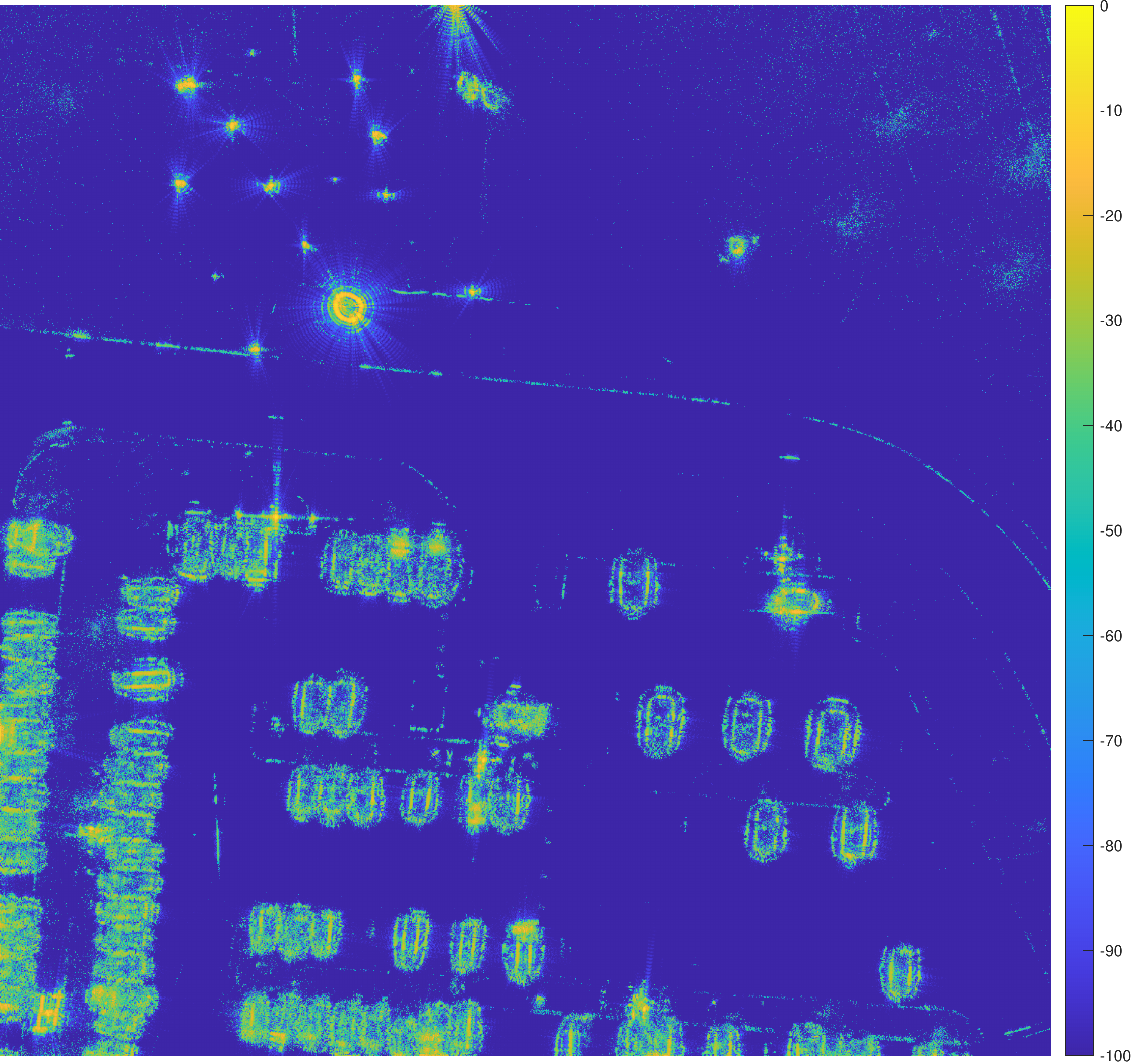}
\includegraphics[width=.3\textwidth]{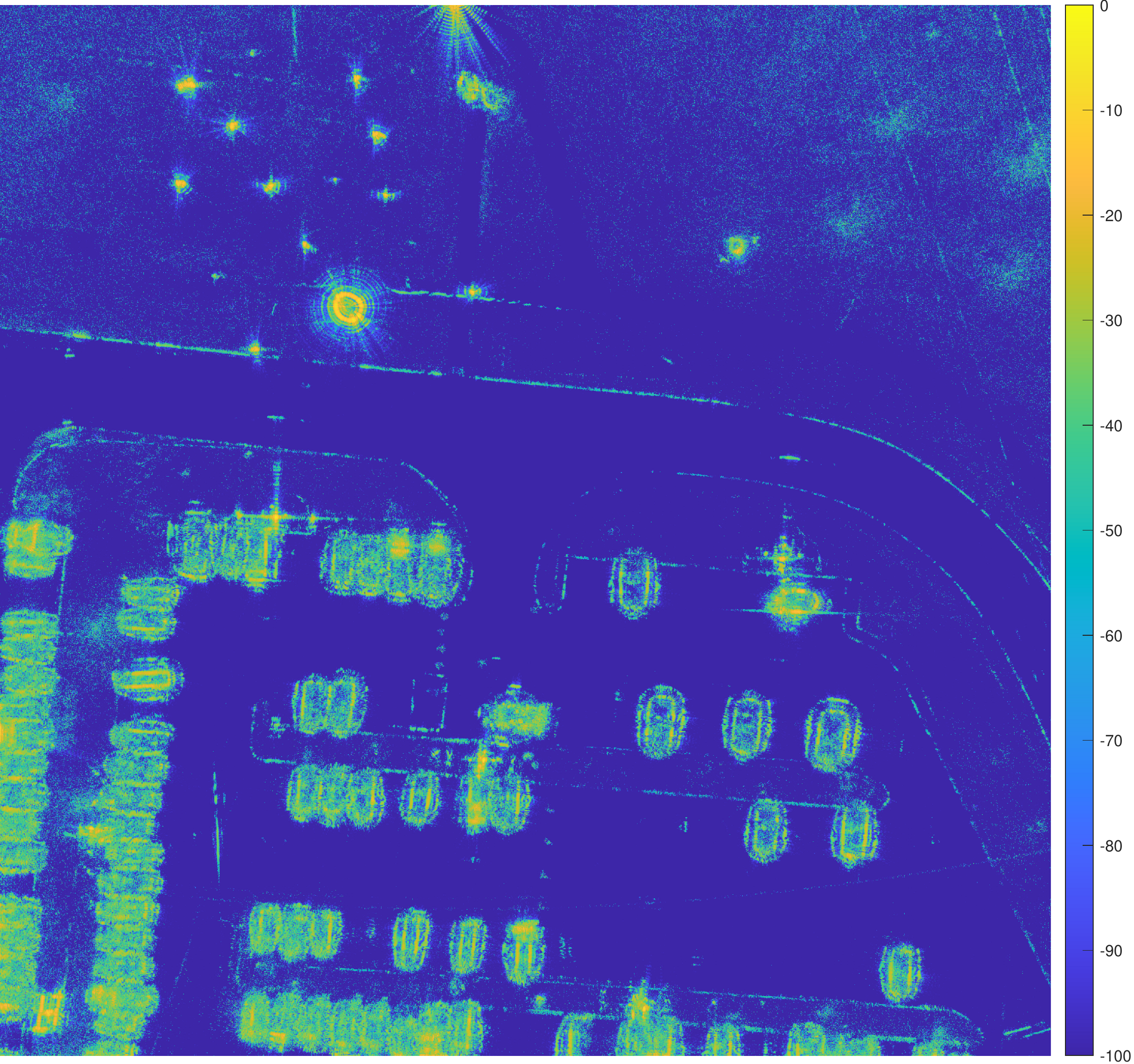}
\caption{Sub-aperture MAP estimate images from the GOTCHA data set, \cite{casteel2007challenge}, formed with the $\ell_1$ method \eqref{eq:map} using $\mathbf{T}=\mathbf{I}$  with $\beta=1$ and $\lambda=1/20,1/40,1/60,1/80$ (left to right, top to bottom).}
\label{fig:GOTCHA_l1}
\end{figure}

An assumption about sparsity is often used to improve image quality. Indeed, there are many sparsity-based SAR image formation methods, see e.g.~\cite{ccetin2014sparsity} for a good overview and \cite{archibald2016image,ccetin2001feature,cetin2005sar,ccetin2014sparsity,potter2010sparsity,potter2008sparse,samadi2011sparse,scarnati2018recent,scarnati2018joint} for specific examples. Because SAR images are frequently sparse in the image domain (more precisely in the magnitude of the image), an $\ell_1$ (or more generally $\ell_p$) norm penalty term on the presumed sparse domain of $\mathbf{f}^{(l)}$ is often added to improve on the results obtained using \eqref{eq:MLestimate}. In this regard it is important to recall that each $\mathbf{f}^{(l)}$ is complex, and only its magnitude, $|\mathbf{f}^{(l)}|$, has a corresponding sparse domain, as the phase is not modeled as sparse, \cite{jakowatz2012spotlight}. Since $|\mathbf{f}^{(l)}|$ is not differentiable, it is convenient to instead employ a unitary diagonal matrix $\Theta^{(l)}$ such that
$(\Theta^{(l)})^\ast\mathbf{f}^{(l)} \approx |\mathbf{f}^{(l)}|$, where $\Theta^{(l)}_{jj} =\mathbf{\tilde{f}}^{(l)}_j/|\mathbf{\tilde{f}}^{(l)}_j|$  is extracted from some cheaply computed approximation $\mathbf{\tilde{f}}^{(l)}$, such as the NUFFT, \cite{sanders2017composite}. Thus we arrive at 
\begin{align}
\mathbf{f}^{(l)}_{\ell_1} &= \arg\min_\mathbf{f} \left \{\frac{\beta^{(l)}}{2} \|\mathbf{\hat{f}}^{(l)}-\mathbf{F}^{(l)}\mathbf{f}\|_2^2+\lambda\|\mathbf{T}(\Theta^{(l)})^\ast\mathbf{f}\|_1\right\},\quad\quad l = 1,\ldots, L,
\label{eq:map}
\end{align}
where $\mathbf{T}$ is an appropriate sparsifying operator.\footnote{In general we can have $\mathbf{T}=\mathbf{T}^{(l)}$, $l=1,\ldots,L$, but choosing sparsifying operators for different sub-apertures may be difficult in practice without prior knowledge of the scene.}  The solution is typically found via the alternating direction method of multipliers (ADMM), \cite{boyd2011distributed}, although other methods are also available. Besides being an inexact representation, approximating $|\mathbf{f}^{(l)}|$ in this way has an important consequence.  In particular, since \eqref{eq:map} regularizes the sparsity of an approximation to the magnitude (instead of the magnitude itself), the regularization term no longer corresponds to any prior distribution since data are considered in forming the initial estimate. An alternate formulation is to solve for the phase explicitly, e.g. \cite{ccetin2014sparsity}. Nevertheless, for practical purposes such empirically based priors are regularly used \cite{zhang2022empirical}. A more commonly discussed difficulty in  using \eqref{eq:map} is that it often requires fine tuning of the sparsity prior parameter $\lambda$ and noise variance $1/\beta^{(l)}$ (often combined into a single parameter). Assuming there is enough information to choose  the regularization parameters, see e.g. \cite{batu2011parameter,VBJS_Speckle}, the $\ell_1$ regularization approach, often referred to as compressive sensing, \cite{candes2006robust}, can be  a highly effective way to compute a point estimate image for SAR. Figure \ref{fig:GOTCHA_l1} shows four realizations of \eqref{eq:map} using $\mathbf{T}=\mathbf{I}$ (reflecting presumed sparsity in the image magnitude) with $\beta=1$ and $\lambda=1/20,1/40,1/60,1/80$. Compared with the NUFFT images, the $\ell_1$ method removes a lot of the ``background'' scattering which may be advantageous.  However, it is also apparent that  the results are quite sensitive to the choice of the regularization parameter. The images are $2048\times2048$ pixels and use $L=12$ windows each spanning $40^o$ with an overlap of $10^o$. Code and parameters (i.e. $\lambda$ and the number of iterations) for this method are taken from \cite{sanders-imaging,sanders2017composite}.

Another observation can be made from the construction of \eqref{eq:map} from \eqref{eq:MLestimate}. Specifically, had the prior probability distribution
\begin{align}\label{eq:laplaceprior}
p(\mathbf{f}^{(l)}|\lambda) \propto \exp\left(-\lambda\|\mathbf{T}|\mathbf{f}^{(l)}|\|_1\right)
\end{align}
been invoked, the resulting posterior would be
\begin{align}\label{eq:posterior1}
\begin{split}
p(\mathbf{f}^{(l)}|\mathbf{\hat{f}}^{(l)},\beta^{(l)},\lambda) & \propto p(\mathbf{\hat{f}}^{(l)}|\mathbf{f}^{(l)},\beta^{(l)})p(\mathbf{f}|\lambda) \\
&\propto \exp\left(-\frac{\beta^{(l)}}{2}\|\mathbf{\hat{f}}^{(l)}-\mathbf{F}^{(l)}\mathbf{f}^{(l)}\|_2^2-\lambda\|\mathbf{T}|\mathbf{f}^{(l)}|\|_1\right),\quad\quad l = 1,\ldots, L.
\end{split}
\end{align}
Observe that maximizing \eqref{eq:posterior1} yields \eqref{eq:map}, and indeed \eqref{eq:map} is known as a maximum a posteriori (MAP) estimate. Of course this is not the only prior distribution that can be used and others would invoke other \emph{a priori} beliefs. This discussion also makes clear that the regularization penalty term within the cost function imposes the \emph{a priori} belief specified in the prior probability distribution, \cite{stuart2010inverse}. Regardless of which prior distribution is chosen, without additional information, the parameters $\lambda$ or $\beta$ would be difficult to choose. Moreover, finding the maximum is generally not the best way to interrogate a posterior.  Finally, the single image reconstruction provides no quantification of the certainty for which the estimate should be trusted. Hence in what follows we take the position that densities should be sought rather than point estimates, and that the parameters of the prior should also be estimated.

\begin{figure}[h]
\centering
\includegraphics[width=.3\textwidth]{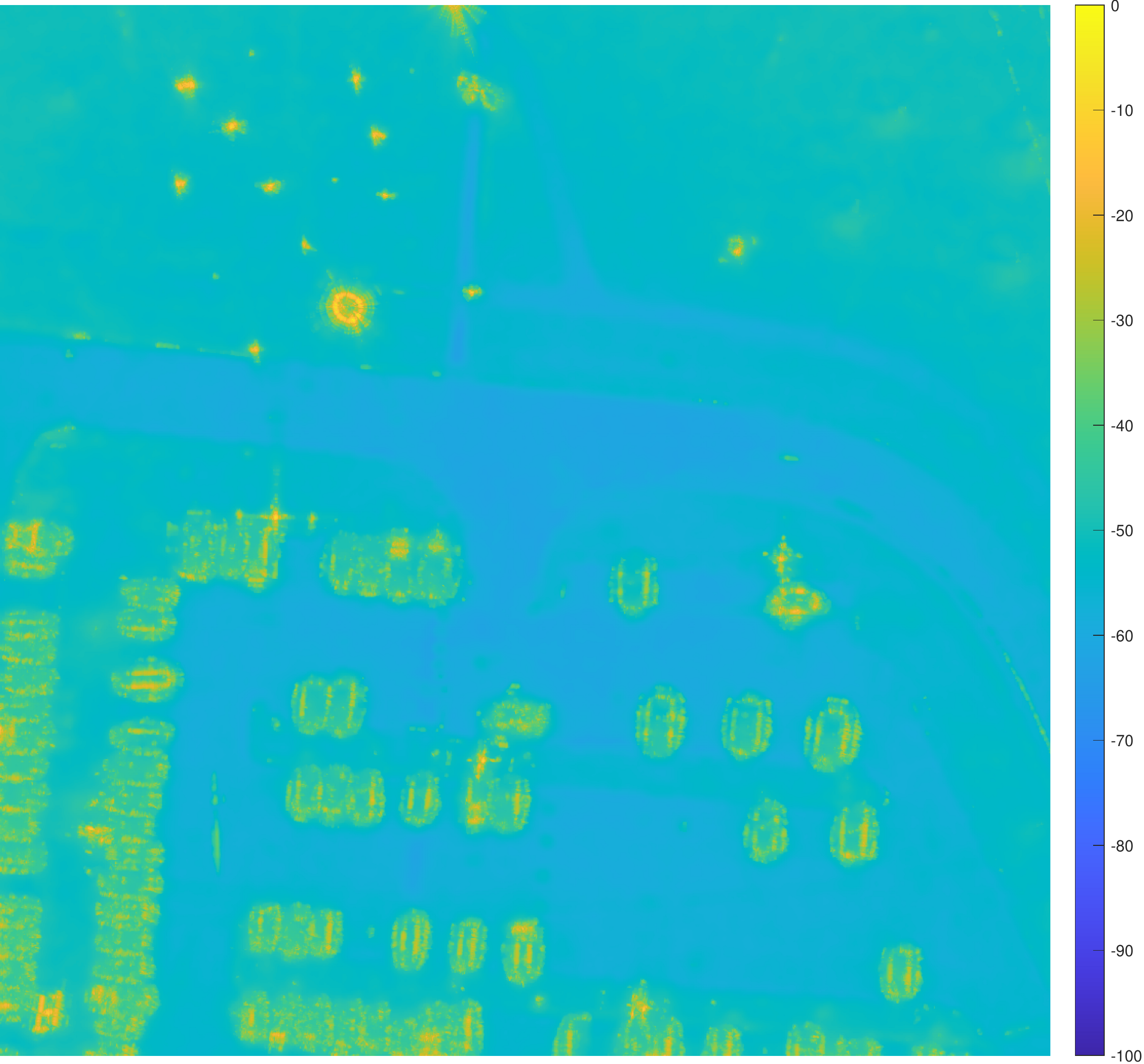}
\includegraphics[width=.3\textwidth]{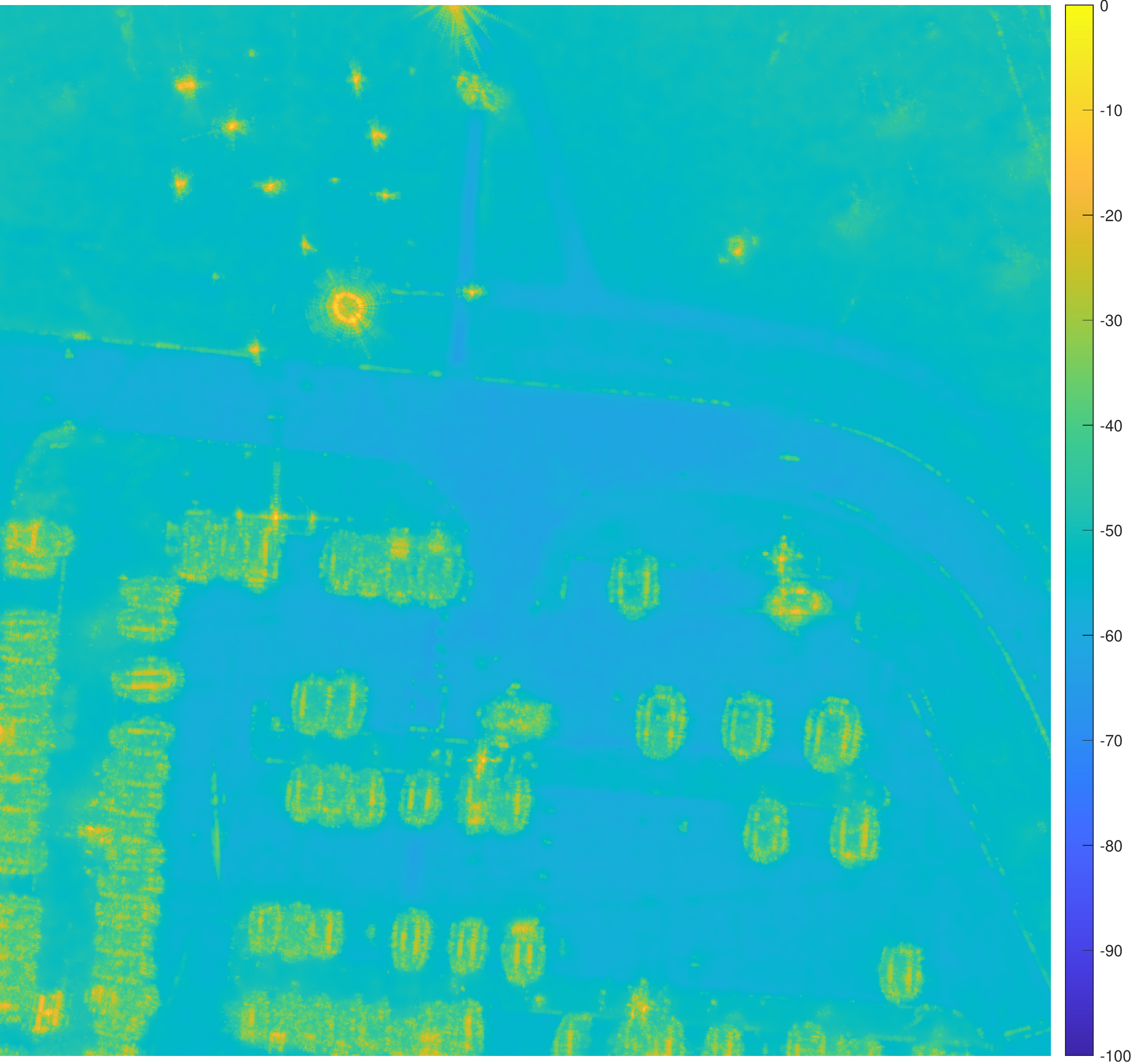}\\
\includegraphics[width=.3\textwidth]{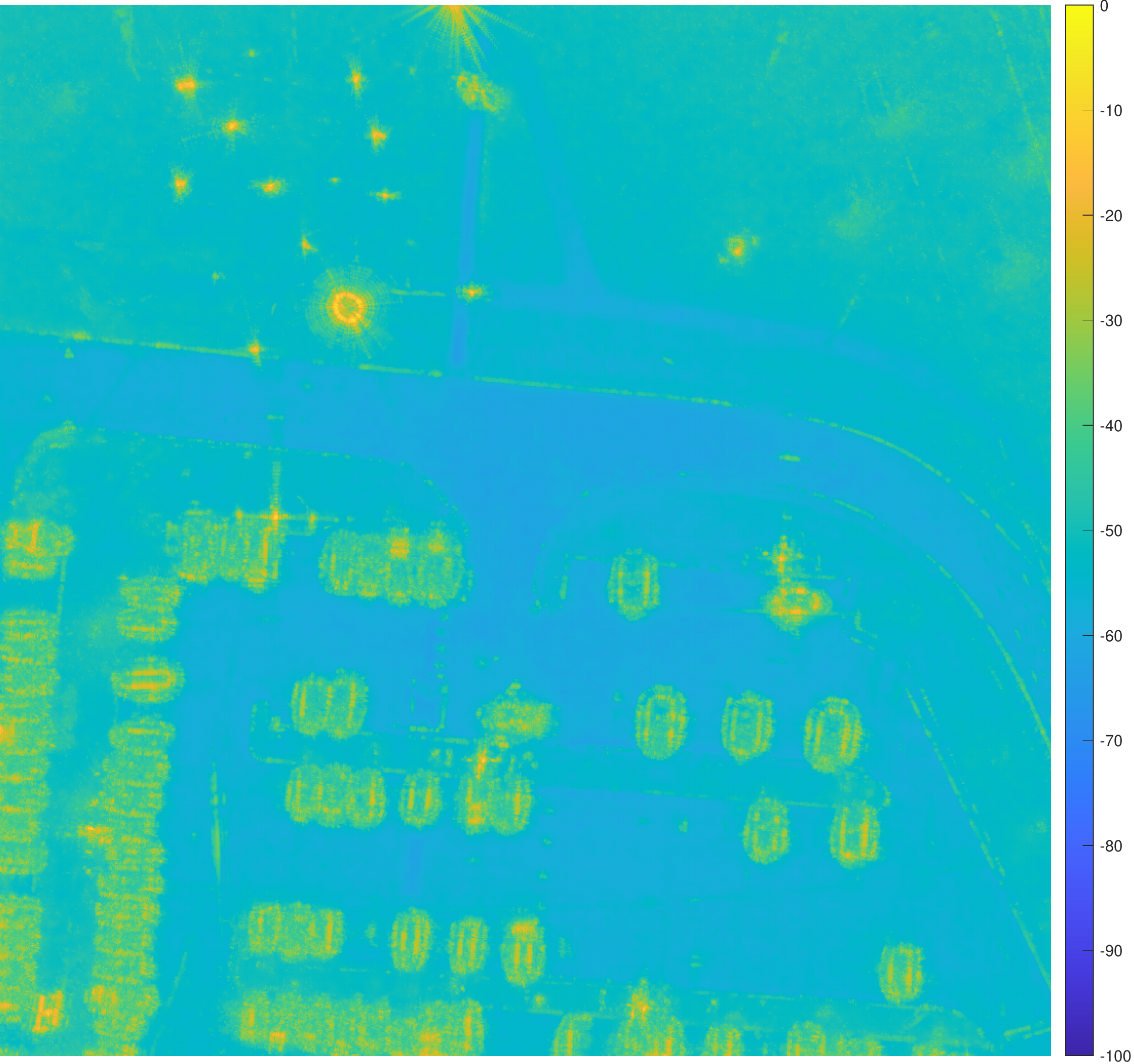}
\includegraphics[width=.3\textwidth]{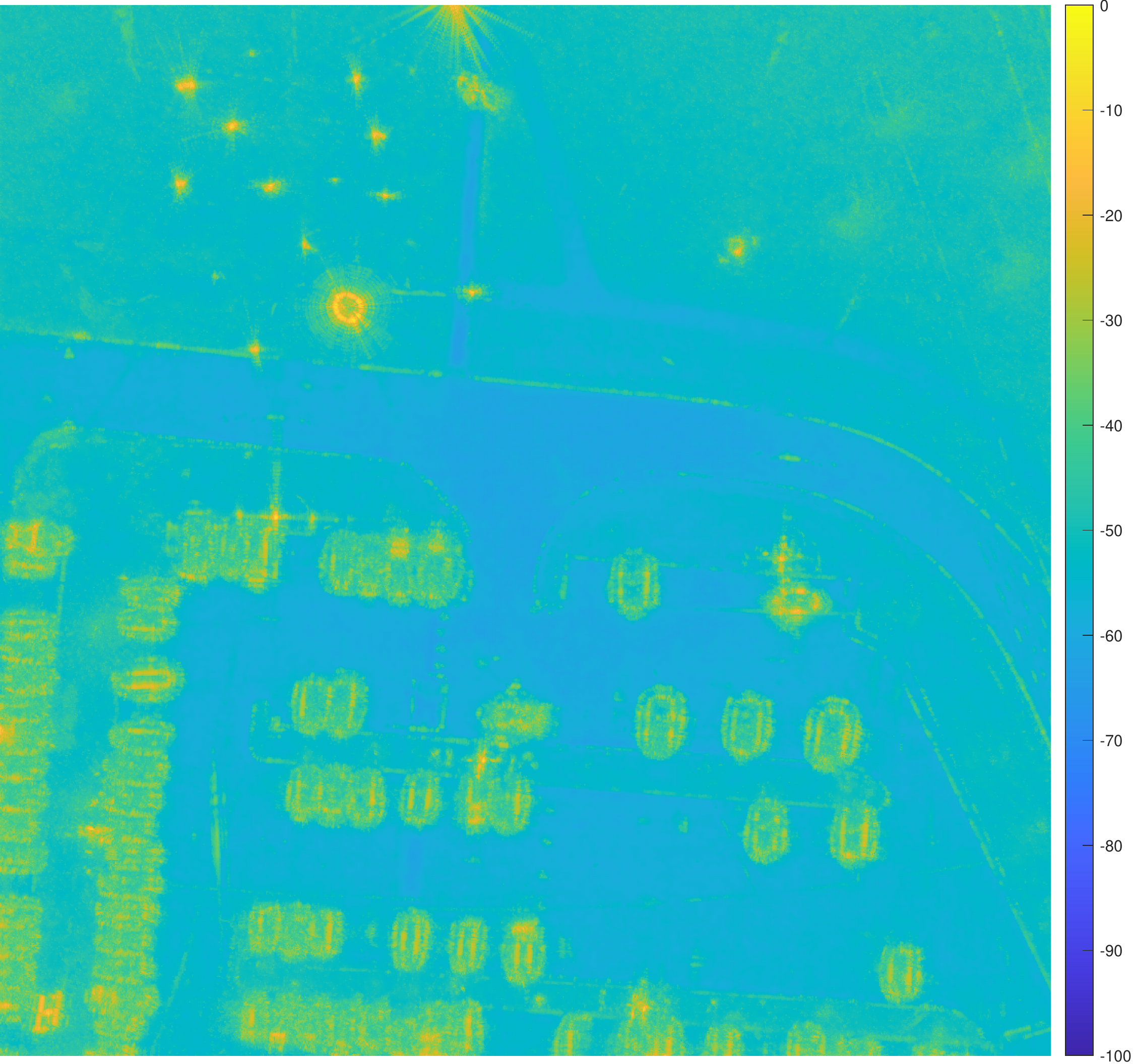}
\caption{Sub-aperture MAP estimate images from the GOTCHA data set, \cite{casteel2007challenge}, formed with the TV-$\ell_1$ method \eqref{eq:map} with $\beta=1$ and $\lambda=1/40,1/80,1/120,1/160$ (left to right, top to bottom).}
\label{fig:GOTCHA_TV}
\end{figure}

Many choices can be made for $\mathbf{T}$, but for comparison purposes, in our experiments we show images formed using two popular sparsity-encouraging image formation methods: (i) $\ell_1$ regularization \cite{tibshirani1996regression}, where sparsity is promoted in the image itself; and (ii) total variation (TV) regularization \cite{rudin1992nonlinear}, where sparsity is promoted in the gradients in the image. Both can be codified as \eqref{eq:map} with sparsifying operator $\mathbf{T}$ chosen appropriately, and both have been extensively applied in SAR (see e.g. \cite{scarnati2018joint,dong2014sar,sanders2017composite,archibald2016image,ccetin2001feature}). Alternatively, the sub-apertures can also be modeled jointly, e.g. \cite{potter2010sparsity,sanders2017composite}, to take advantage of the presumably small differences between neighboring sub-aperture images. Figure \ref{fig:GOTCHA_TV} shows four realizations of \eqref{eq:map} using the two-dimensional anisotropic TV operator for $\mathbf{T}$ with $\beta=1$ and $\lambda=1/40,1/80,1/120,1/160$. Compared with the NUFFT and $\ell_1$ images, we notice the piecewise constant smoothing effect of TV regularization which, depending on $\lambda$, appears to blur some objects in the scene. The images are $1024\times1024$ pixels and use $L=12$ windows each spanning $40^o$ with an overlap of $10^o$. We note that the smaller image size is due to long runtime when using a non-identity regularization operator. Again, code and parameters (i.e. $\lambda$ and the number of iterations) for this method are taken from \cite{sanders-imaging,sanders2017composite}.

\subsection{Combining sub-aperture images}
As mentioned briefly in the descriptions of the figures, if desired, sub-aperture images can be combined to form a single composite image. One way to achieve this is by taking the argument of maximum modulus over the sub-apertures for each image pixel, \cite{moses2004wide}. Specificaly, we compute elementwise
\begin{align}\label{eq:GLRT}
\mathbf{f}^{composite}_{max} = \arg\max_{l=1,\ldots,L} |\mathbf{f}^{(l)}|,
\end{align}
which can be interpreted as a Generalized Likelihood Ratio Test (GLRT) statistic for the scattering responses. Although non-coherent and subject to a reduction of information, this combination of sub-apertures provides a single image to examine and display, and furthermore is able to somewhat mollify the problem of the faulty isoptropic scattering assumption.

\section{Sub-aperture SAR Image Formation with UQ}
\label{sec:compositeSAR}

We now derive the proposed method for sub-aperture SAR imaging with uncertainty quantification. We begin by specifying the hierarchical Bayesian model followed by the SBL-type estimation procedure.

\subsection{Hierarchical prior for speckle and sparsity}\label{sec:specklemodel}

With the likelihood given by \eqref{eq:likelihood}, a prior for each latent variable, $\mathbf{f}^{(l)}$, $l = 1,\ldots,L$, is required to compute a posterior. The prior expresses a belief about a quantity before observation.  For example, as already noted above an $\ell_1$ prior can be used to enforce a sparsity assumption on the magnitude of $\mathbf{f}^{(l)}$ in some domain.  Here we encode into the prior the fact that SAR images are affected by the speckle phenomenon. As discussed in Section \ref{sec:introduction},  speckle, which is manifested as a granular pattern of bright and dark spots thorugh an image,  occurs in all coherent imaging and is often  mischaracterized as noise, \cite{jakowatz2012spotlight}. Although it is in fact signal, speckle still decreases image contrast and so it is desirable to remove it. Speckle reduction is often addressed using denoising techniques such as total variation (TV) regularization, which reduces to the form given by \eqref{eq:map}. But as shown in \cite{CGspeckle} and will be further demonstrated in our numerical experiments, treating speckle as noise results in an unnatural smoothing of the speckle.  Therefore, here we instead follow \cite{CGspeckle,dong2014sar,jakowatz2012spotlight} by directly incorporating the speckle into the prior, so that it is properly characterized as part of the signal.

In fully-developed speckle, we assume $\text{Re}(\mathbf{f}^{(l)}_i)$ and $\text{Im}(\mathbf{f}^{(l)}_i)$, i.e. the real and imaginary parts of the $i$th pixel of the image $\mathbf{f}^{(l)}$, are i.i.d.~Gaussian with variance $1/\boldsymbol{\alpha}^{(l)}_{i}$. Hence $\mathbf{f}^{(l)}$ is circularly-symmetric complex Gaussian, i.e. $\mathbf{f}^{(l)}\sim\mathcal{CN}(\mathbf{0},\text{diag}(1/\boldsymbol{\alpha}^{(l)}))$, with density
\begin{align}\label{eq:jointprior}
p(\mathbf{f}^{(l)}|\boldsymbol{\alpha}^{(l)}) \propto \prod_{i=1}^{N} \boldsymbol{\alpha}^{(l)}_i \exp\left(-\frac12\|\sqrt{\boldsymbol{\alpha}^{(l)}}\odot\mathbf{f}^{(l)}\|_2^2\right), \quad\quad l = 1,\ldots, L,
\end{align}
where $\odot$ is elementwise multiplication. Thus the prior on the magnitude of the $i$th pixel $|\mathbf{f}^{(l)}_i|=\sqrt{\text{Re}(\mathbf{f}^{(l)}_i)^2+\text{Im}(\mathbf{f}^{(l)}_i)^2}$ is Rayleigh with mean proportional to $1/\boldsymbol{\alpha}^{(l)}_i$. 

\begin{remark}
({\em Multiplicative noise model}). Because any change in the magnitude of each pixel $|\mathbf{f}^{(l)}_i|$ is proportional to $1/\boldsymbol{\alpha}^{(l)}_{i}$, the speckle phenomenon has also been modeled as a multiplicative noise, \cite{dong2014sar}. By contrast, here we address the speckle directly by including it in our model with the prior \eqref{eq:jointprior}, and later simultaneously estimating the associated speckle parameter $1/\boldsymbol{\alpha}^{(l)}$ rather than using post-image-formation techniques.
\end{remark}

While the prior \eqref{eq:jointprior} and likelihood \eqref{eq:likelihood} are indeed enough to derive each  posterior $p(\mathbf{f}^{(l)}|\mathbf{\hat{f}}^{(l)},\boldsymbol{\alpha}^{(l)},\beta^{(l)})$ and compute a MAP estimate for each sub-aperture, this estimate would not necessarily be representative of such a posterior. Nor does it provide information about the statistical confidence of the estimate for each recovered pixel value, or for any other recovered features of the image, \cite{nagy2002image}. Moreover, the regularization parameters for both the cost function and prior in the MAP estimate approach (analogous to $\beta^{(l)}$ and $\boldsymbol{\alpha}^{(l)}$ here) are {\em user-specified}.  However they are truly unknown and therefore should be inferred. 
For these reasons we seek the joint posterior $p(\mathbf{f}^{(l)},\boldsymbol{\alpha}^{(l)},\beta^{(l)} | \mathbf{\hat{f}}^{(l)})$, hence we will be simultaneously estimating the speckle parameter $\boldsymbol{\alpha}^{(l)}$ and noise parameter $\beta^{(l)}$, which will lend clarity when determining whether or not the speckle reduction techniques are actually working.\footnote{Without a reference ground truth image, speckle statistics are typically only estimated from small regions of already formed images, \cite{argenti2013tutorial}.} 

To calculate $p(\mathbf{f}^{(l)},\boldsymbol{\alpha}^{(l)},\beta^{(l)} | \mathbf{\hat{f}}^{(l)})$, we need to define priors on $\boldsymbol{\alpha}^{(l)}$ and $\beta^{(l)}$. We first invoke a conjugate Gamma prior for $\beta^{(l)}$. That is, $\beta^{(l)}\sim\Gamma(c,d)$ with density
\begin{align}\label{eq:betaprior}
p(\beta^{(l)}|c,d) \propto (\beta^{(l)})^{c-1}\exp(-d\beta^{(l)}), \quad\quad l = 1,\ldots, L.
\end{align}
Similarly, a conjugate Gamma prior is invoked on each element of $\boldsymbol{\alpha}^{(l)}$, i.e. $\boldsymbol{\alpha}^{(l)}_{i}\sim\Gamma(a,b)$. By independence, $\boldsymbol{\alpha}^{(l)}\sim\Gamma(a,b)$ with
\begin{align}\label{eq:priorA}
p(\boldsymbol{\alpha}^{(l)}|a,b) \propto & \prod_{i=1}^{N} (\boldsymbol{\alpha}^{(l)})^{a-1}_i \exp\left(-b\sum_{i=1}^{N}\boldsymbol{\alpha}^{(l)}_i\right), \quad\quad l = 1,\ldots,L.
\end{align}
Note the dependence of \eqref{eq:betaprior} and \eqref{eq:priorA} on parameters $a$, $b$, $c$, and $d$, which as in \cite{bardsley2012mcmc,tipping2001sparse} are chosen rather than inferred. Following \cite{tipping2001sparse}, we choose $c = d = 0$ so that the resulting improper prior on the noise variance \eqref{eq:betaprior} is uniform over a logarithmic scale, making all scales equally likely.  Regarding $a$ and $b$,  analogous parameters for a real-valued model in \cite{bardsley2012mcmc} are chosen to reflect the uncertainty in the latent variable, making the prior uninformative. On the other hand in \cite{CGspeckle,tipping2001sparse}, $a =b  =0$, resulting in an improper prior $p(\mathbf{f}^{(l)}_i)\sim1/|\mathbf{f}^{(l)}_i|$, which is peaked at zero and hence encourages sparsity.\footnote{To ensure numerical robustness, we choose all parameters, $a,b,c,d$ to be machine precision rather than $0$ in our implementation.} As already noted,  the reflectivity in SAR images is presumably sparse, so that using \eqref{eq:priorA} with parameters $a, b \approx 0$  seems reasonable.  Importantly, fixing $a, b, c$ and $d$ in this way removes any need for user-defined parameters in this model.

The form of the posterior is achieved through a hierarchical Bayesian model where the likelihood parameters $\mathbf{f}^{(l)}$ and $\beta^{(l)}$ are given priors with prior parameters (hyperparameters) $\boldsymbol{\alpha}^{(l)}$, $c$, and $d$. Moving up to the final level of hierarchy in this model, the hyperparameter $\boldsymbol{\alpha}^{(l)}$ is given a prior (called a hyperprior) with hyperhyperparameters $a$ and $b$. By Bayes' theorem, the joint posterior for each $\mathbf{f}^{(l)}$, $\boldsymbol{\alpha}^{(l)}$, and $\beta^{(l)}$, $l = 1,\ldots,L$,  is
\begin{align}\label{eq:posterior}
p(\mathbf{f}^{(l)},\boldsymbol{\alpha}^{(l)},\beta^{(l)}|\mathbf{\hat{f}}^{(l)}) &\propto p(\mathbf{\hat{f}}^{(l)}|\mathbf{f}^{(l)},\beta^{(l)})p(\beta^{(l)}|0,0)p(\mathbf{f}^{(l)}|\boldsymbol{\alpha}^{(l)})p(\boldsymbol{\alpha}^{(l)}|0,0) \\
&\propto (\beta^{(l)})^{M-1}\exp\left(-\frac{\beta^{(l)}}{2}\|\mathbf{\hat{f}}^{(l)}-\mathbf{F}^{(l)}\mathbf{f}^{(l)}\|_2^2-\frac12\|\sqrt{\boldsymbol{\alpha}^{(l)}}\odot\mathbf{f}^{(l)}\|_2^2\right). \nonumber
\end{align}
It is important to make the distinction between the speckle model introduced in \cite{dong2014sar} and the subsequent methodology developed in \cite{CGspeckle}. While the implementation in \cite{dong2014sar} provides an appropriate characterization of speckle, it then uses TV regularization to obtain a MAP estimate. By contrast, the characterization in \cite{CGspeckle} is followed by a full posterior recovery using a sampling method which enables quantification of the uncertainty.  Note that the methods in both \cite{dong2014sar} and \cite{CGspeckle} are for full azimuth imaging ($L = 1$ in \eqref{eq:posterior}).

\subsubsection{Regularization in other domains}

Applying the above sparsifying prior in domains other than the imaging domain is challenging. Some methods have been proposed that accommodate specific domains, e.g. TV \cite{babacan2010sparse,chantas2008variational,chantas2009variational,chantas2006bayesian,churchill2020estimation}, however a generalized regularization model, for which the only restriction was that the regularization operator satisfy the  so-called common kernel condition (which will be discussed in more detail in Section \ref{sec:estimation_new}), \cite{kaipio2006statistical},  was not considered until \cite{glaubitz2022generalized}.\footnote{{Incidentally, the method introduced in \cite{glaubitz2022generalized}  restricts the noise to be independent but not necessarily identically distributed, as is required in other approaches.  This desirable property may be useful for fusing multiple time-dependent SAR data sets, but is not part of the current investigation.}}

Hence, while \eqref{eq:posterior} properly models speckle, as mentioned earlier one may wish to enforce sparsity in some other domain, e.g. some transformation of the magnitude $\mathbf{T}|\mathbf{f}^{(l)}|$. In order to include this, one can alter \eqref{eq:jointprior} to
\begin{align}\label{eq:jointpriorT}
p(\mathbf{f}^{(l)}|\boldsymbol{\alpha}^{(l)}) \propto \prod_{i=1}^{N} \boldsymbol{\alpha}^{(l)}_i \exp\left(-\frac12\|\sqrt{\boldsymbol{\alpha}^{(l)}}\odot (\mathbf{T}|\mathbf{f}^{(l)}|)\|_2^2\right),\quad\quad l = 1,\ldots,L.
\end{align}
Letting $\Theta^{(l)}$ be a unitary diagonal matrix as in Section \ref{sec:estimate} such that $(\Theta^{(l)})^\ast\mathbf{f}^{(l)} \approx |\mathbf{f}^{(l)}|$,  %\footnote{$\Theta^{(l)$ can also be estimated though right? E.g. updated at each step by the same formula using the new $\mathbf{f}$? Is this what you guys did before? Then we can just include it in the posterior.} \AG{Yes, $\Theta$ can be separated out and estimated.  We did not have a SAR example in the paper with Jan, but the BCD algorithm is able to upate each variable. We did update $\Theta$ in the autofocusin paper (with Theresa).}}.
then we have
\begin{align}\label{eq:jointpriorTTheta}
p(\mathbf{f}^{(l)}|\boldsymbol{\alpha}^{(l)},\Theta^{(l)}) \propto \prod_{i=1}^{N} \boldsymbol{\alpha}^{(l)}_i \exp\left(-\frac12\|\sqrt{\boldsymbol{\alpha}^{(l)}}\odot (\mathbf{T}(\Theta^{(l)})^\ast \mathbf{f}^{(l)})\|_2^2\right).
\end{align}
We can then write the final posterior for each $l = 1,\ldots, L$ as
\begin{align}
p(\mathbf{f}^{(l)},\boldsymbol{\alpha}^{(l)},\beta^{(l)}|\mathbf{\hat{f}}^{(l)},\Theta^{(l)}) &\propto p(\mathbf{\hat{f}}^{(l)}|\mathbf{f}^{(l)},\beta^{(l)})p(\beta^{(l)}|0,0)p(\mathbf{f}^{(l)}|\boldsymbol{\alpha}^{(l)},\Theta^{(l)})p(\boldsymbol{\alpha}^{(l)}|0,0) \\
&\propto (\beta^{(l)})^{M-1}\exp\left(-\frac{\beta^{(l)}}{2}\|\mathbf{\hat{f}}^{(l)}-\mathbf{F}^{(l)}\mathbf{f}^{(l)}\|_2^2-\frac12\|\sqrt{\boldsymbol{\alpha}^{(l)}}\odot(\mathbf{T}(\Theta^{(l)})^\ast \mathbf{f}^{(l)})\|_2^2\right). \nonumber
\end{align}
%\HELP{Here is where Victor will put in the update for $\Theta$ following our discussion.}

\subsection{Estimation Techniques Revisited}
\label{sec:estimation_new}
%\AG{I reorganized this paragraph a bit.}
At this stage the proposed method deviates from \cite{CGspeckle}, where $\mathbf{f}^{(l)}$, $\boldsymbol{\alpha}^{(l)}$, and $\beta^{(l)}$ were sampled from the joint posterior using a single $360^\circ$ azimuth window ($L = 1$).   Requirements for storage and memory of samples for $\mathbf{f}^{(l)}$, $\boldsymbol{\alpha}^{(l)}$, and $\beta^{(l)}$ were already challenging in the single aperture window case, limiting image reconstruction to images of $N=512^2$ and smaller.  In particular, there were $\sim5\times10^5$ sampled parameters to consider, each with $\sim1300$ samples.  In the current framework, as a consequence of appropriately modeling the anisotropic nature of scatterers in the scene, the composite image reconstruction model in \eqref{eq:discretemodel} is now roughly $L$ times larger than problem size considered in the corresponding models in \cite{CGspeckle}. Specifically in the context of \eqref{eq:posterior},  we would essentially be reconstructing $L$ sub-aperture images of size $N$.     
For example, the corresponding composite imaging problem with $L=60$ would require us to store samples of $\sim3\times10^{7}$ parameters.  

To limit such extreme storage and memory requirements, we now instead follow a Bayesian coordinate descent (BCD) algorithm similar to that of generalized sparse Bayesian learning method \cite{glaubitz2022generalized} to deterministically choose a density estimate. SBL-type methods have been used extensively for reconstructing SAR images from phase history data, see e.g. \cite{liu2013superresolution,xu2012bayesian,xue2009sar}. In addition to these direct applications of SBL in SAR image reconstruction from phase history data, the posterior density arrived at in SBL has also been used in sampling schemes for a variety of SAR tasks such as moving target recognition \cite{newstadt2014moving}, model selection for speckle \cite{karakucs2018generalized}, as well as both direct image reconstruction \cite{CGspeckle} and composite image reconstruction \cite{wu2015high}. To form an image, the individual (conditional) posterior for $\mathbf{f}^{(l)}$ and to the parameters $\boldsymbol{\alpha}^{(l)}$ and $\beta^{(l)}$, $l = 1,\ldots,L,$ are updated in sequence. The individual posterior for each $\mathbf{f}^{(l)}$ is 
\begin{align}
p(\mathbf{f}^{(l)}|\mathbf{\hat{f}}^{(l)},\boldsymbol{\alpha}^{(l)},\beta^{(l)},\Theta^{(l)}) &\propto \exp\left(-\frac{\beta^{(l)}}{2}\|\mathbf{\hat{f}}^{(l)}-\mathbf{F}^{(l)}\mathbf{f}^{(l)}\|_2^2-\frac12\|\sqrt{\boldsymbol{\alpha}^{(l)}}\odot(\mathbf{T}(\Theta^{(l)})^\ast \mathbf{f}^{(l)})\|_2^2\right),
\end{align}
hence
\begin{align}
\mathbf{f}^{(l)}|\mathbf{\hat{f}}^{(l)},\boldsymbol{\alpha}^{(l)},\beta^{(l)},\Theta^{(l)} \sim \mathcal{CN}(\boldsymbol{\mu}^{(l)},\boldsymbol{\Sigma}^{(l)})
\end{align}
with
\begin{align}\label{eq:bayesf}
\boldsymbol{\mu}^{(l)} &= \beta^{(l)}\boldsymbol{\Sigma}^{(l)}(\mathbf{F}^{(l)})^\ast\mathbf{\hat{f}}^{(l)}
\end{align}
and
\begin{align}\label{eq:bayesf2}
\boldsymbol{\Sigma}^{(l)} &= \left[\beta^{(l)}(\mathbf{F}^{(l)})^\ast\mathbf{F}^{(l)}+(\mathbf{T}\Theta^{(l)})^\ast\text{diag}(\boldsymbol{\alpha}^{(l)})\mathbf{T}\Theta^{(l)}\right]^{-1}.
\end{align}
%\HELP{Victor will want to add the update to $\Theta$ above.}
We note that for this to hold, $\mathbf{F}^{(l}$ and $\mathbf{T}\Theta^{(l)}$ must satisfy the common kernel condition \cite{glaubitz2022generalized,kaipio2006statistical,tikhonov1995numerical}: $$\text{kernel}(\mathbf{F}^{(l)})\cap\text{kernel}(\mathbf{T}\Theta^{(l)})=\{\mathbf{0}\}, \quad l=1,\ldots,L,$$ which of course depends on the choice of $\mathbf{T}$, which can be non-invertible and even non-square. Since $\mathbf{F}^{(l)}$ is a Fourier transform operator with a full-rank matrix representation then it has trivial kernel. Therefore, for the purposes of our examples, we confirmed via direct computation that both $\mathbf{T}$ being the identity and the anisotropic TV operator when multiplied by $\Theta^{(l)}$ yield matrices with trivial kernels.

Per Bayesian coordinate descent \cite{glaubitz2022generalized}, $\boldsymbol{\alpha}^{(l)}$ and $\beta^{(l)}$, $l = 1,\ldots,L$,  are approximated  by the means of their conditional posteriors. Due to the conjugate prior relationship, both conditional posteriors are Gamma-distributed %$$\beta^{(l)}|\cdots \sim \Gamma(M,\|\mathbf{\hat{f}}^{(l)}- \mathbf{F}^{(l)}\mathbf{f}^{(l)}\|^2)$$ and $$\boldsymbol{\alpha}^{(l)}|\cdots \sim \Gamma(1,\|\mathbf{f}^{(l)}\|^2.$$ Therefore, the updates are just the means of these distributions. For $\beta^{(l)}$, it's
with means given by
\begin{align}\label{eq:betamean}
\mu_{\beta^{(l)}} = \frac{M}{\|\mathbf{\hat{f}}^{(l)}- \mathbf{F}^{(l)}\mathbf{f}^{(l)}\|_2^2}
\end{align}
and
\begin{align}\label{eq:alphamean}
\mu_{\boldsymbol{\alpha}^{(l)}} = \frac{1}{\|\mathbf{T}(\Theta^{(l)})^\ast\mathbf{f}^{(l)}\|_2^2}.
\end{align}

\begin{remark}
Note that while $\Theta^{(l)}$ is not included in the estimation problem as an inferred variable, it is not fixed by an initial estimate as in Section \ref{sec:estimate} e.g. as $\Theta^{(l)}_{jj} = (\mathbf{f}_{ML})_j^{(l)}/|(\mathbf{f}_{ML})_j^{(l)}|$. In the iterative estimation procedure that follows we update each $\Theta^{(l)}$ by the same formula instead using the current update for $\mathbf{f}^{(l)}$:
\begin{align}\label{eq:Theta}
\Theta^{(l)}_{jj} = \mathbf{f}^{(l)}_j/|\mathbf{f}^{(l)}_j|, \quad j=1,\ldots,N, \quad l = 1,\ldots,L.
\end{align}
This technique was used in \cite{scarnati2018joint} to keep the quantity $(\Theta^{(l)})^*\mathbf{f}^{(l)}$ real-valued as the iterates change. Observe that due to construction of the prior this has no impact for the case when $\mathbf{T}=\mathbf{I}$.
\end{remark}

The algorithm therefore consists of alternating computation of $\boldsymbol{\mu}$ and $\boldsymbol{\Sigma}$ from \eqref{eq:bayesf} and \eqref{eq:bayesf2}, $\boldsymbol{\alpha}$ and $\beta$ from \eqref{eq:betamean} and \eqref{eq:alphamean}, and $\Theta^{(l)}$ from \eqref{eq:Theta}, until a convergence criterion has been achieved. The computational challenge is in applying the covariance $\boldsymbol{\Sigma}$ to compute $\boldsymbol{\mu}$ at each step. However, as mentioned earlier, since $\mathbf{F}$ is applied via a NUFFT, \cite{fessler2003nonuniform}, if $\mathbf{T}=\mathbf{I}$ (the assumption generally used for the SAR speckle model)  then $\boldsymbol{\Sigma}^{-1}=\beta\mathbf{F}^H\mathbf{F}+(\mathbf{T}\Theta^{(l)})^\ast\text{diag}(\boldsymbol{\alpha})\mathbf{T}\Theta^{(l)}$ is diagonal and hence efficiently inverted with a cheap elementwise division by $\beta+\boldsymbol{\alpha}$, \cite{CGspeckle}. If $\mathbf{T}$ is more general, then solving the linear system is more involved. Nevertheless, in many cases $\boldsymbol{\Sigma}^{-1}$ is typically sparse, e.g. if $\mathbf{T}$ is a TV operator. Algorithm \ref{alg:bcd} shows the exact steps used in our examples.

\begin{algorithm}[h]
\caption{Sub-aperture SAR image formation}
\label{alg:bcd}
\begin{algorithmic}

\STATE{Initiate $\boldsymbol{\mu}^{(l)}_0=(\mathbf{F}^{(l)})^\ast\mathbf{\hat{f}}^{(l)}$, $\Theta^{(l)}_0 = \boldsymbol{\mu}^{(l)}_0/|\boldsymbol{\mu}_0^{(l)}|$, and convergence tolerance $\epsilon$.}

\FOR{$l=1,\ldots,L$}

\STATE{$k = 0$;}

\WHILE{$\frac{\|\boldsymbol{\mu}_{k+1}^{(l)}-\boldsymbol{\mu}_{k}^{(l)}\|_2}{\|\boldsymbol{\mu}_{k}^{(l)}\|_2} >\epsilon$ }

\STATE{$\boldsymbol{\alpha}_{k+1}^{(l)} = \frac{1+2a}{|\mathbf{T}(\Theta^{(l)})^\ast\boldsymbol{\mu}_k^{(l)}|^2+2b}$;}

\STATE{$\beta_{k+1}^{(l)}=\frac{M+2c}{\|\mathbf{\hat{f}}^{(l)}-\mathbf{F}^{(l)}\boldsymbol{\mu}^{(l)}_k\|_2^2+2d}$;}

\STATE{$\boldsymbol{\mu}^{(l)}_{k+1} =\left[\beta^{(l)}_{k+1}(\mathbf{F}^{(l)})^\ast\mathbf{F}^{(l)}+(\mathbf{T}\Theta^{(l)})^\ast\text{diag}(\boldsymbol{\alpha}_{k+1}^{(l)})\mathbf{T}\Theta^{(l)}\right]^{-1}\beta_{k+1}^{(l)}\boldsymbol{\mu}^{(l)}_0$;}

\STATE {$\Theta^{(l)}_{k+1} = \boldsymbol{\mu}_{k+1}^{(l)}/|\boldsymbol{\mu}_{k+1}^{(l)}|$;}

\STATE{$k=k+1$;}

\ENDWHILE

\STATE{$\boldsymbol{\alpha}^{(l)} = \boldsymbol{\alpha}^{(l)}_{k+1}$;}
\STATE{$\beta^{(l)} = \beta^{(l)}_{k+1}$;}
\STATE{$\boldsymbol{\mu}^{(l)} = \boldsymbol{\mu}^{(l)}_{k+1}$;}
\STATE{$\Theta^{(l)} = \Theta^{(l)}_{k+1}$;}

\ENDFOR

\end{algorithmic}
\end{algorithm}

Many typical convergence criteria can be used. In our implementation, we base convergence on the relative change of subsequent iterates (see details in Section \ref{sec:results}), although convergence could also be based on speckle reduction in a particular region based on $\boldsymbol{\alpha}$, etc. Moreover, the sub-aperture images can be formed in parallel resulting in a roughly $L$ times acceleration (corresponding to the for loop in Algorithm \ref{alg:bcd}).

The results of Algorithm \ref{alg:bcd} are $L$ probability distributions for each of the sub-aperture SAR images, where as specified in \eqref{eq:bayesf} and \eqref{eq:bayesf2}, the mean and covariance of these complex Gaussians are defined by the final iterates $\{\boldsymbol{\alpha}^{(l)},\beta^{(l)},\boldsymbol{\mu}^{(l)},\Theta^{(l)}\}_{l=1}^L$. Being probability distributions, many estimates can be derived from these distributions, or they can be sampled to form other estimates, etc. This provides more information than is common in sub-aperture wide angle SAR image reconstruction. In addition, note that $\boldsymbol{\alpha}^{(l)}$ and $\beta^{(l)}$ are also themselves point estimates of the speckle and noise parameters, respectively. This is significant as estimates for such parameters (if possible at all) would require additional processing as they are not included in the image formation process  itself. Moreover, as discussed in more detail in \cite{CGspeckle} with respect to the $L = 1$ (full azimuth) case, in general we have no intuition for $\boldsymbol{\alpha}^{(l)}$ and $\beta^{(l)}$.  Our method allows us to encode this uncertainty by choosing uninformative priors.  As a result, the estimates we obtain help to lend clarity when determining whether or not the speckle (and noise) reduction techniques are actually working. Finally, without a ground truth reference image, any additional processing to obtain this information would have to rely on pre-formed images. 

\subsection{Combining sub-aperture densities}

As mentioned above in \eqref{eq:GLRT} and originating in \cite{moses2004wide}, taking the argument of the  maximum modulus over the sub-apertures of each image pixel provides a way to form a single composite image from multiple sub-aperture image estimates. This typically brightens the appearance of the image compared with full-azimuth imaging (see Figure \ref{fig:GOTCHA_widecomp}), but is non-coherent and a reduction of the information in each sub-aperture image estimate. Furthermore, it is applied to image estimates, which are already a reduction of information compared to the full posterior density.

We therefore propose a method for combining the sub-aperture \emph{densities} themselves. This option assumes that each sub-aperture image is independent of the others and takes advantage of the properties of summing independent Gaussian random variables. We have that each sub-aperture image is modeled by a Gaussian posterior with mean $\boldsymbol{\mu}^{(l)}$ and covariance $\boldsymbol{\Sigma}^{(l)}$, which are computed using the final estimates for the noise variance and speckle parameter.  If we assume each of these random variables are independent\footnote{We recognize that given the typical processing that uses overlapping data sub-apertures, azimuthal angular independence may not typically be a good assumption. However, we suggest that non-overlapping sub-apertures can be taken. Furthermore, any angular independence argument would also depend on the scene itself.}, then their average is also a Gaussian random variable. That is, the coherent composite posterior is modeled as
\begin{align}
\mathbf{f}^{composite}_{mean}\sim\mathcal{CN}(\boldsymbol{\mu}^{composite},\boldsymbol{\Sigma}^{composite}),
\end{align}
where
\begin{align}
\boldsymbol{\mu}^{composite} = \frac1L \sum_{l=1}^L\boldsymbol{\mu}^{(l)}
\end{align}
and
\begin{align}
\boldsymbol{\Sigma}^{composite} = \frac{1}{L^2}\sum_{l=1}^L \boldsymbol{\Sigma}^{(l)}.
\end{align}
There are many ways to interrogate this composite distribution. For example, taking the elementwise square root of the diagonal of $\boldsymbol{\Sigma}^{composite}$ gives the standard deviation image for the composite posterior, which gives insight beyond a single point estimate into the spread of the density. For instance, around $68\%$ of the mass lies within a standard deviation of the mean.

We furthermore note that the significance of the availability to estimate the composite speckle parameter
\begin{align}
\boldsymbol{\alpha}^{composite} = \frac1L \sum_{l=1}^L \boldsymbol{\alpha}^{(l)}
\end{align}
which similar to the variance can provide confirmation that speckle has been reduced. Finally, of course it is also possible to use the maximum option to recover a non-coherent point estimate while relying on the mean option to obtain other information.

\begin{figure}[h!]
\centering
\includegraphics[width=.8\textwidth]{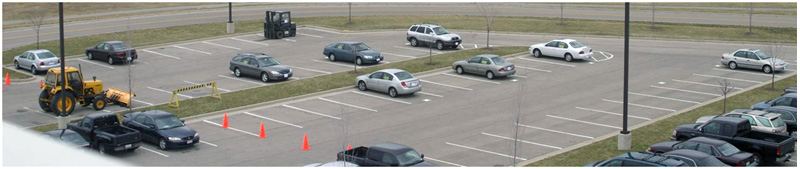}\\
\includegraphics[width=.4\textwidth]{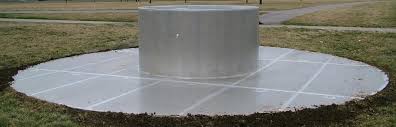}
\includegraphics[width=.4\textwidth]{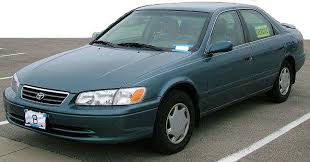}
\caption{Optical images of parking lot being imaged in GOTCHA dataset, \cite{casteel2007challenge}. Note scene contains a variety of calibration targets, such as primitive reflectors like the tophat shown, a Toyota Camry, forklift, and tractor.}
\label{fig:parking_lot}
\end{figure}

\section{Computational Results}\label{sec:results}

\begin{figure}[h]
\centering
\includegraphics[width=.3\textwidth]{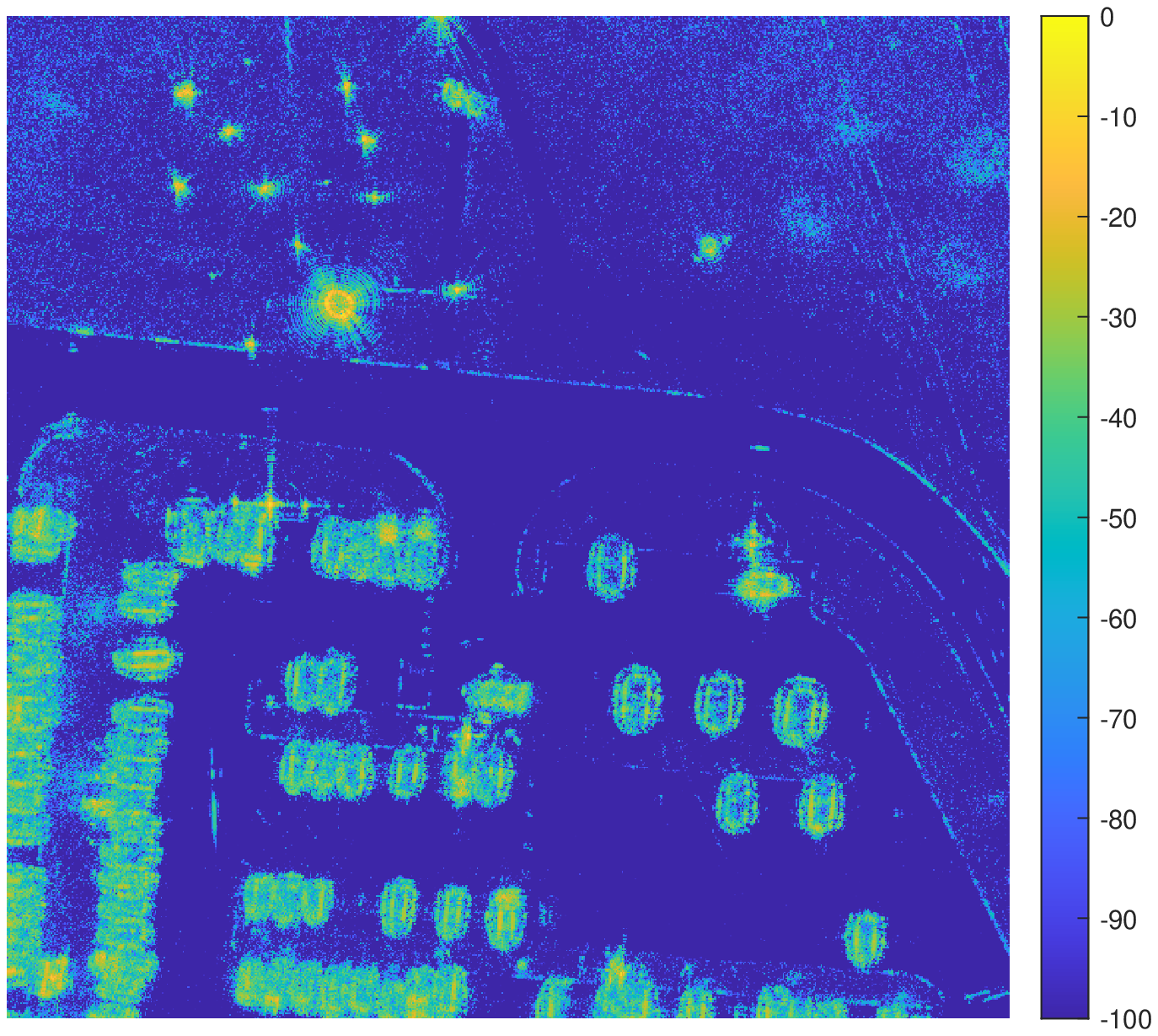}
\includegraphics[width=.3\textwidth]{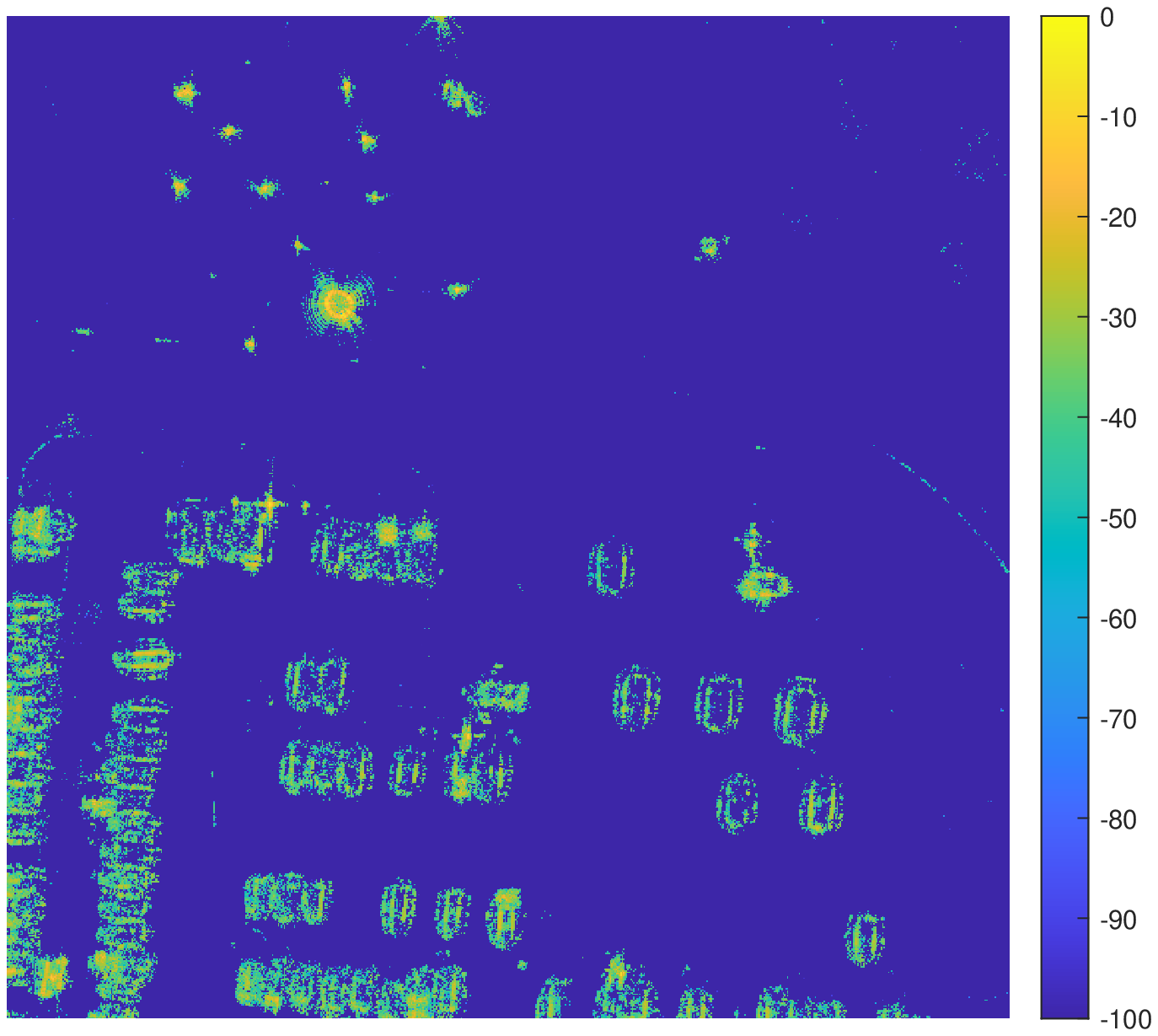}\\

\caption{Sub-aperture composite mean images formed with Algorithm \ref{alg:bcd} using $\mathbf{T}=\mathbf{I}$ and convergence tolerance $\epsilon=0.1$ (left) and $\epsilon=0.01$ (right) from the GOTCHA data set, \cite{casteel2007challenge}.}
\label{fig:GOTCHA_mean}
\end{figure}

\begin{figure}[h]
\centering
\includegraphics[width=.3\textwidth]{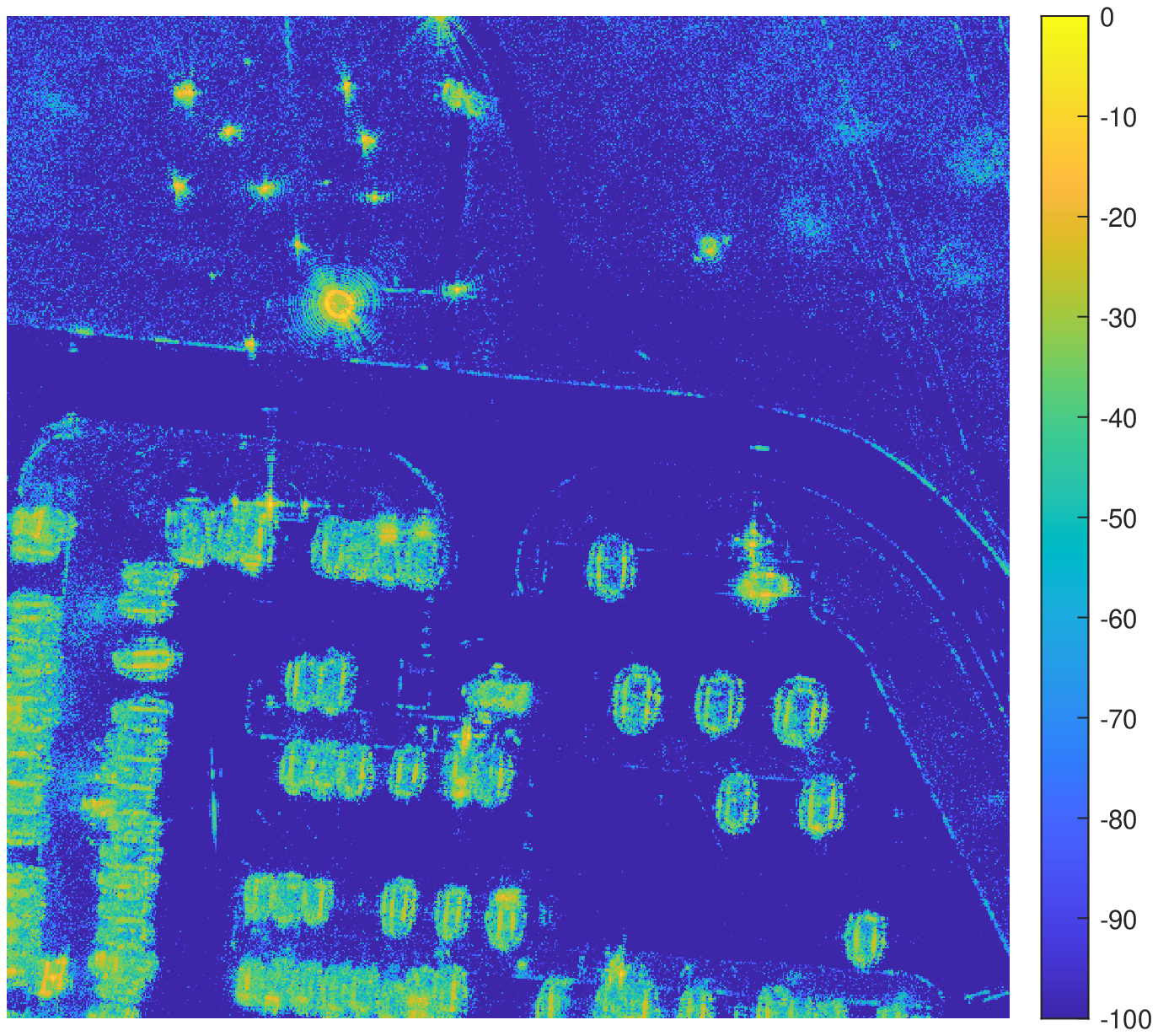}
\includegraphics[width=.3\textwidth]{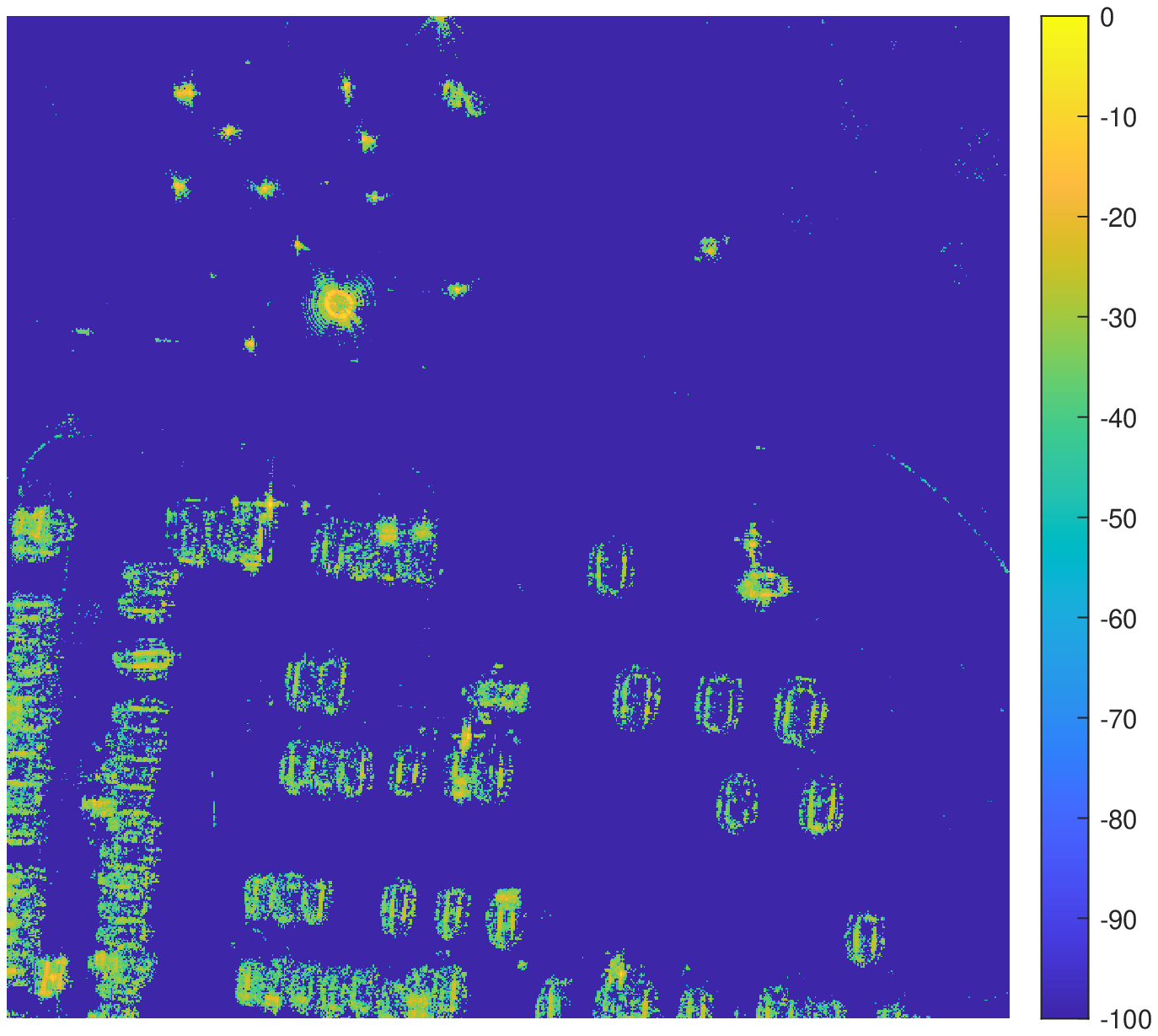}\caption{Sub-aperture composite max images formed with Algorithm \ref{alg:bcd} using $\mathbf{T}=\mathbf{I}$ and convergence tolerance $\epsilon=0.1$ (left) and $\epsilon=0.01$ (right) from the GOTCHA data set, \cite{casteel2007challenge}.}
\label{fig:GOTCHA_max}
\end{figure}

\subsection{GOTCHA SAR Data}
The GOTCHA Volumetric SAR Data Set 1.0 consists of SAR phase history data of a parking lot scene collected at X-band with a 640 MHz bandwidth with full azimuth coverage at 8 different elevation angles with full polarization, \cite{casteel2007challenge}. The GOTCHA SAR phase histroy data were collected when a plane carrying a sensor flew a roughly circular measurement flight around a parking lot near the Sensors Directorate Building at Wright-Patterson Air Force Base in Dayton, Ohio.  The parking lot contains various targets including civilian vehicles, construction vehicles, calibration targets, primitive reflectors, and military vehicles. Figure \ref{fig:parking_lot} shows optical images of the targets. Note that because this is real-world data, the elevation angle is not perfectly constant, and the path is not perfectly circular.
The center frequency is 9.6GHz and bandwidth is 640MHz. This public release data has been used extensively for testing new SAR image formation methods, \cite{austin2012sparse,austin2009sparse,ertin2007gotcha,sanders2017composite},  and is available from \textbf{https://www.sdms.afrl.af.mil/index.php?collection=gotcha}.

\subsection{Experiments}

We now demonstrate the accuracy, efficiency, and robustness of the proposed method for sub-aperture SAR image reconstruction from phase history data with uncertainty quantification. Note that the ground truth reflectivity image is unknown, preventing the computation of standard error statistics such as the relative error. This is the case even in synthetically-created SAR examples, where the true reflectivity is still unknown. Therefore, the uncertainty quantification information the proposed method provides is all the more valuable, as it is able to quantify how much we should trust pixel values and structures in the image even in the absence of ground truth. Throughout, all reflectivity images $\mathbf{f}$ are displayed in decibels (dB):
\begin{align}
20\log_{10}\left(\frac{|\mathbf{f}|}{\max |\mathbf{f}|}\right),
\end{align}
with a minimum of $-100$ dB and maximum of $0$ dB. Lesser or greater values are assigned the minimum or maximum.

In what follows, we generally hesitate to make subjective claims about the appearance or ``quality'' of these reconstructions, as all certainly have advantages and disadvantages. Thus depending on what the image is being used for, different image reconstruction methods or image features (e.g. smoothness, sparsity, etc.) may be more useful. Instead, we focus on the new methodology, its capabilities, as well as objective comparisons.

\subsubsection{Benchmarking sparsity testing}

Figures \ref{fig:GOTCHA_mean} and \ref{fig:GOTCHA_max} show the composite mean and max estimates for the GOTCHA parking lot scene using $L=12$ windows each spanning $40^o$ with $10^o$ of overlap. As demonstrated in Figure \ref{fig:GOTCHA_mean} compared with other sub-aperture methods in Figures \ref{fig:GOTCHA_widecomp} (right), \ref{fig:GOTCHA_l1}, and \ref{fig:GOTCHA_TV}, Algorithm \ref{alg:bcd} is able to reduce background speckle appearance while localizing bright targets.

We note that different convergence tolerances yield different results. E.g., if $\epsilon=0.1$ (corresponding to a $10\%$ relative error iterate threshold), the sub-aperture regions each run for $\sim 3$ iterations and still have visible speckle in them, evidenced by the speckle parameter image. The image contains all features of the original image with increased contrast. However, if $\epsilon=0.01$ (corresponding to a $1\%$ relative error iterate threshold), the sub-aperture regions each run for $\sim12$ iterations and have very little visible speckle in them. Strong targets are localized more precisely, but the image is less interpretable to human eyes.

\begin{figure}[h]
\centering
\includegraphics[width=.3\textwidth]{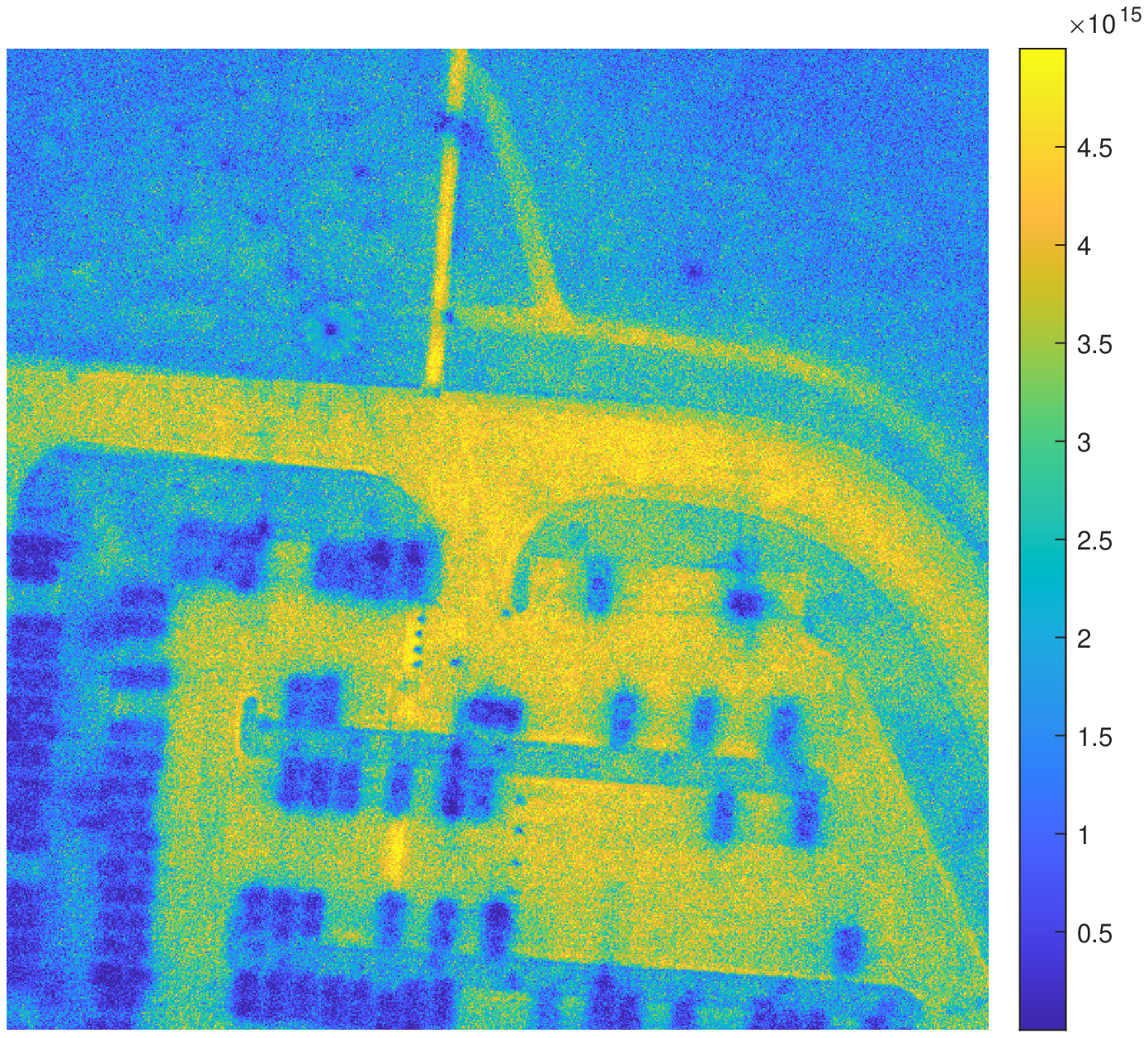}
\includegraphics[width=.3\textwidth]{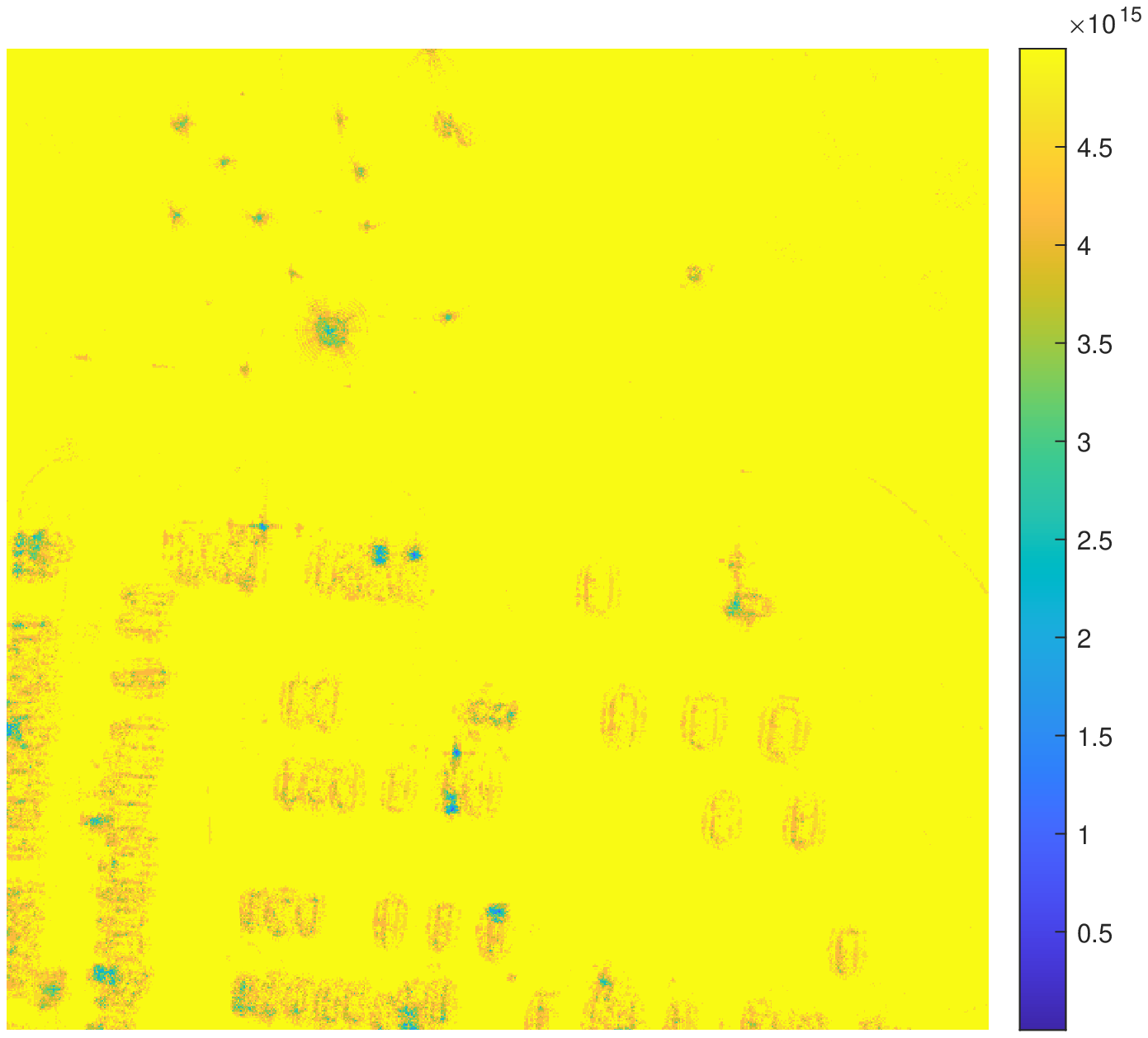}
\caption{Sub-aperture composite speckle parameter images formed with Algorithm \ref{alg:bcd} using $\mathbf{T}=\mathbf{I}$ and convergence tolerance $\epsilon=0.1$ (left) and $\epsilon=0.01$ (right) from the GOTCHA data set, \cite{casteel2007challenge}.}
\label{fig:GOTCHA_sp}
\end{figure}

\begin{figure}[h]
\centering
\includegraphics[width=.3\textwidth]{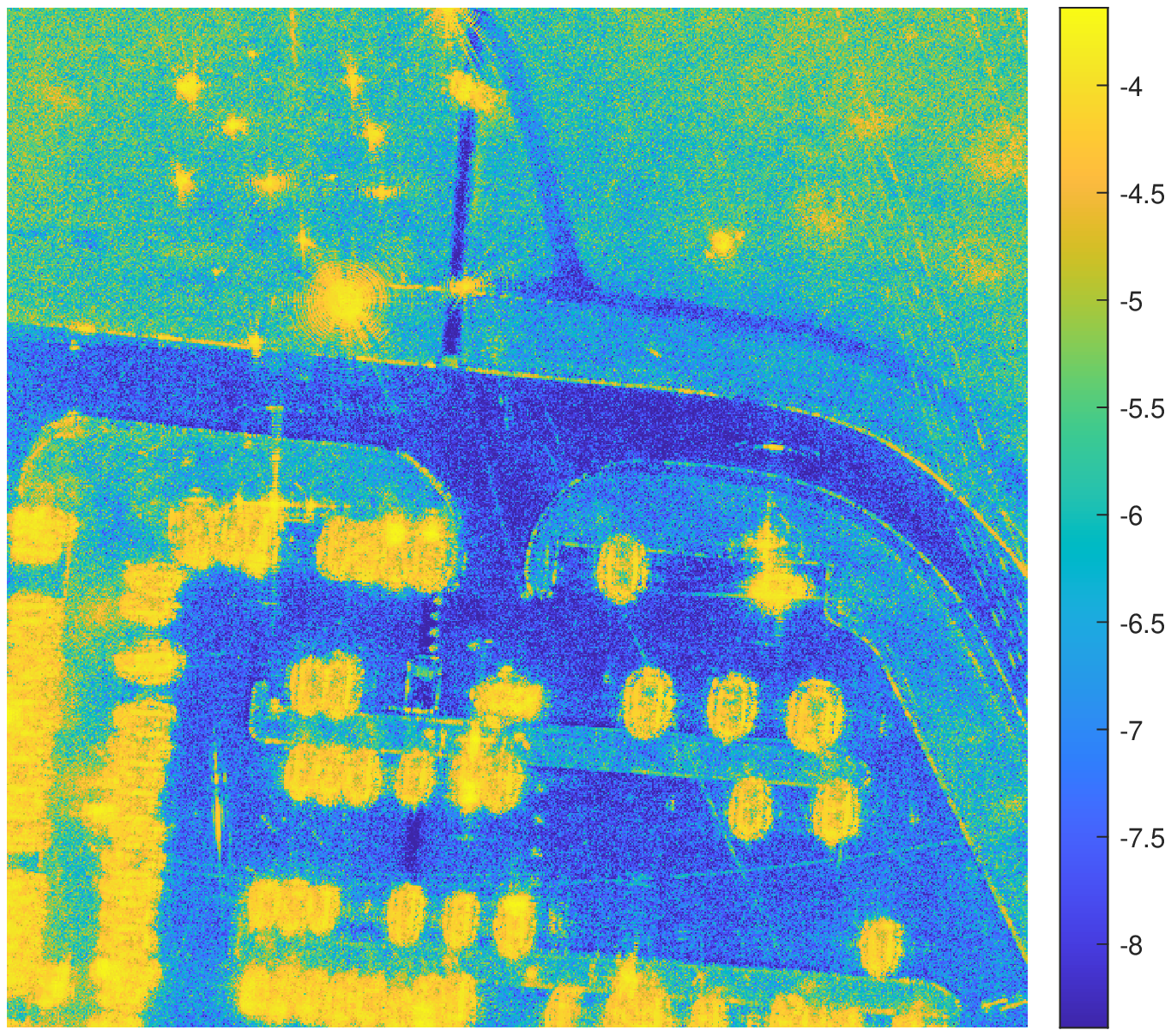}
\includegraphics[width=.3\textwidth]{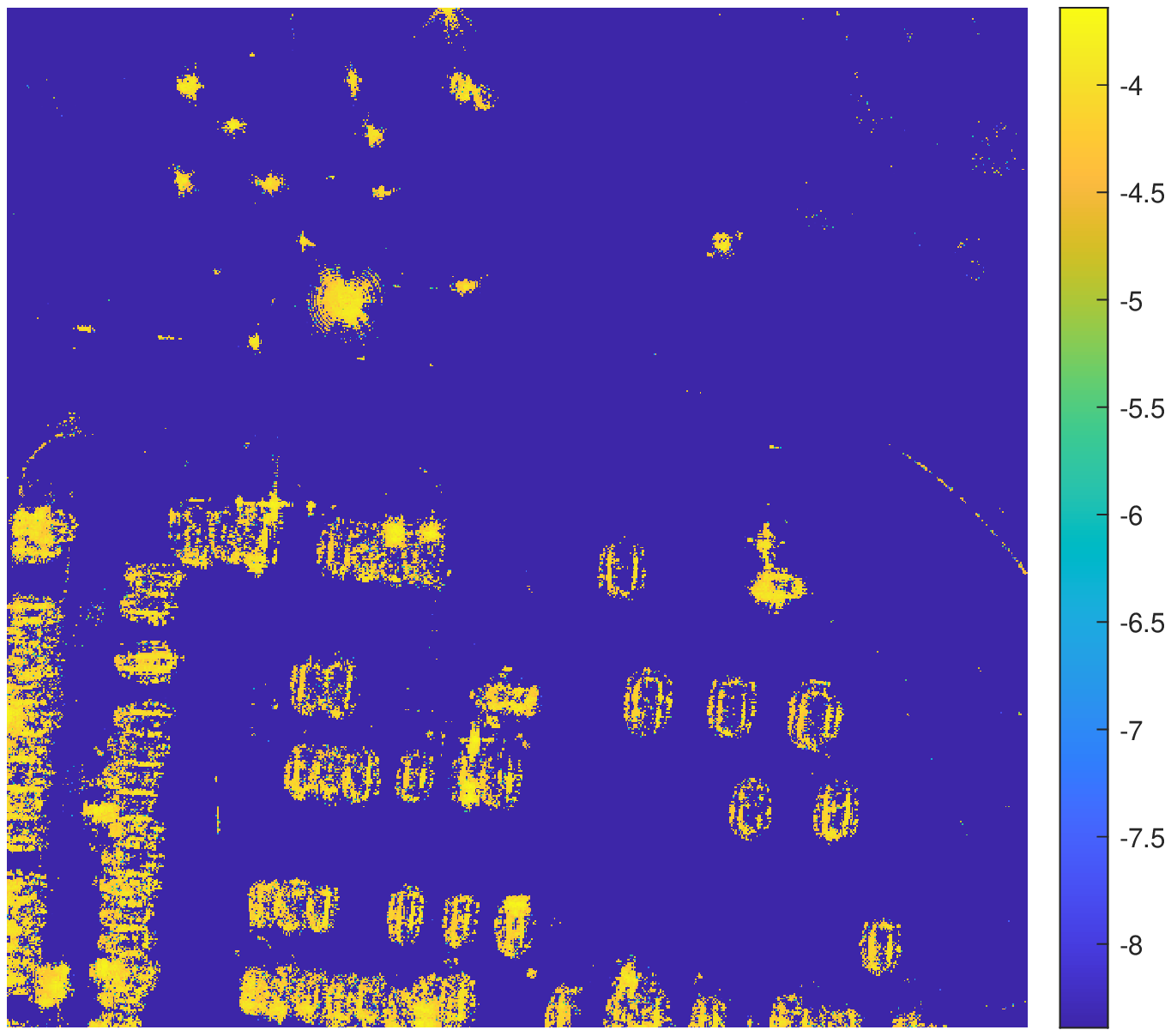}
\caption{Sub-aperture composite standard deviation images formed with Algorithm \ref{alg:bcd} using $\mathbf{T}=\mathbf{I}$ and convergence tolerance $\epsilon=0.1$ (left) and $\epsilon=0.01$ (right) from the GOTCHA data set, \cite{casteel2007challenge}.}
\label{fig:GOTCHA_std}
\end{figure}

Furthermore, the proposed method also supplies extra information in Figures \ref{fig:GOTCHA_sp} and \ref{fig:GOTCHA_std}, which show the speckle parameter and standard deviation images, respectively. These images confirm the speckle and noise reduction resulting from the sparsity-inducing hierarchical Bayesian prior, indicating tight confidence intervals in the background with the only significant variance occurring where the mean estimate predicted signal returns.

\begin{table}[h]
\begin{center}
\begin{tabular}{c|c}
	\hline
Method & Variance\\
\hline
NUFFT & $5.5275e-08$ \\
$\ell_1$, $\lambda=1/20$ & $2.0548e-16$ \\
Alg. 1, T=I, $\epsilon=.1$, mean & $2.4433e-18$ \\
Alg. 1, T=I, $\epsilon=.1$, max & $3.5544e-16$ \\
Alg. 1, T=I, $\epsilon=.01$, mean & $2.1432e-28$ \\
Alg. 1, T=I, $\epsilon=.01$, max & $1.0437e-26$ \\
\hline
\end{tabular}
\caption{Variance for a small $50\times50$-pixel homogeneous subregion with each algorithm to show speckle reduction in $2048\times2048$ images.}
\label{table:variance}
\end{center}
\end{table}

To quantify the improvement and speckle reduction, Table \ref{table:variance} shows the variance of each image in a small ($50 \times 50$ pixel) homogeneous region containing no targets in the center of the image in the intersection entering the parking lot. We choose to compare the most sparsifying $\ell_1$ example tested ($\lambda=1/20$), as this will serve as a lower variance estimate compared with other regularization parameter choices.
%Due to long runtime for the large $2048\times2048$, we omit the TV-$\ell_1$ result.
This type of measurement has been used to evaluate speckle reduction, \cite{argenti2013tutorial}. For the result, we see that all methods perform quite well, with all but one Algorithm 1 composite resulting in smaller variance compared with the most sparsifying $\ell_1$ method over the homogeneous subregion, indicating superior small scale speckle reduction.

%Upon closer inspection we further observe the ``blocky'' nature of the TV regularized image. This is consistent with choosing the gradient domain to be sparse.  More specifically, the speckle is treated as noise (rather than as signal) and smoothed out into piecewise constant structures.

\begin{figure}[h]
\centering
\includegraphics[width=\textwidth]{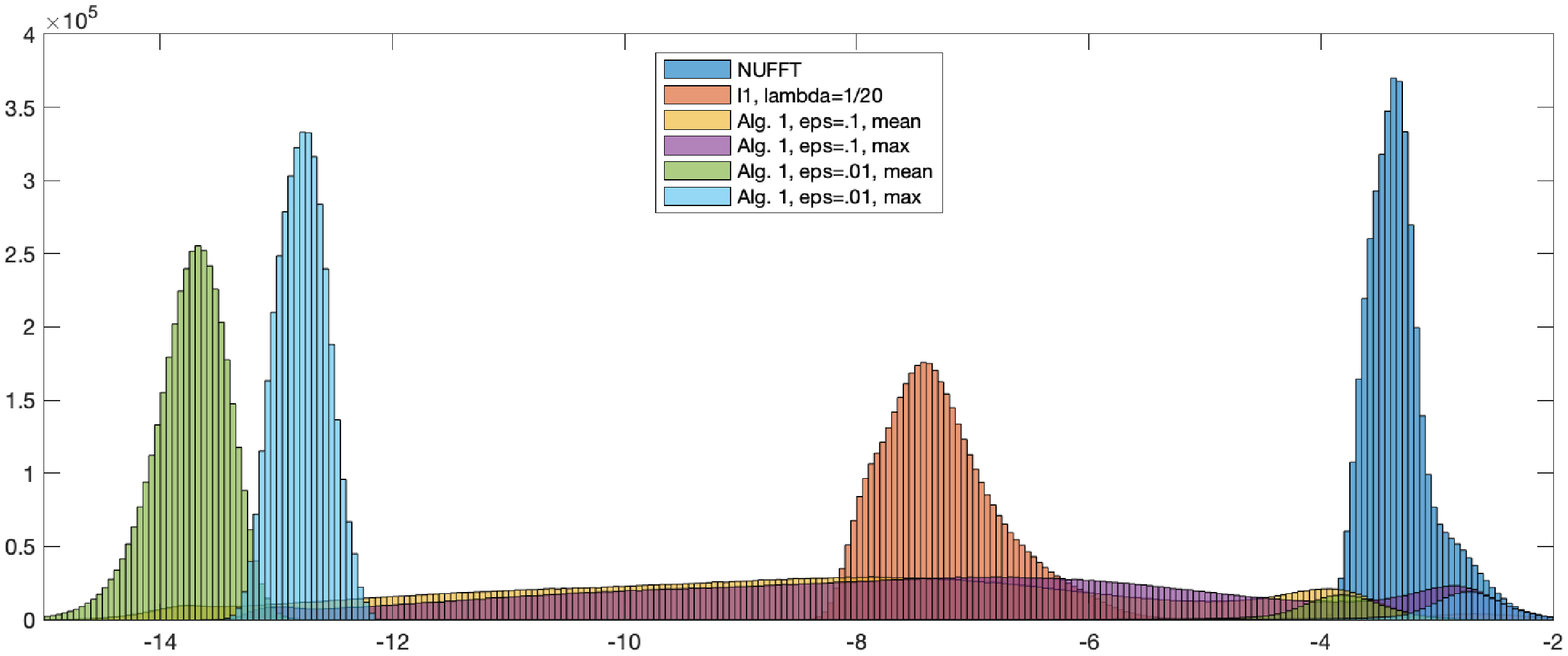}
\caption{Histograms of $\log_{10}{|\mu|}$, where $\mu$ is the composite mean and max resulting from Algorithm \ref{alg:bcd} with different convergence tolerances, and the point estimates of the NUFFT and $\ell_1$ regularization with $\lambda=1/20$ comparing their sparsification. Observe that Algorithm \ref{alg:bcd} sparsifies the image more and has visibly increased contrast.}
\label{fig:contrast}
\end{figure}

\begin{table}[h]
\begin{center}
\begin{tabular}{c|cc}
	\hline
	image size & $2048^2$ \\
	\hline
	\textbf{NUFFT}  & 8.4s \\
	\textbf{$\ell_1$ (20 iters.)} & 2534.7s \\
%	\textbf{TV (20 iters.)} & 16165s & X \\
	\textbf{Alg. 1 (T=I, $\epsilon=0.1$)} & 21.3s \\
	\textbf{Alg. 1 (T=I, $\epsilon=0.01$)} & 94.2s \\
%	\textbf{Alg. 1 (R=TV, $\epsilon=0.1$)} & & \\
%	\textbf{Alg. 1 (R=TV, $\epsilon=0.01$)} & & \\
	\hline
\end{tabular}
\caption{Sub-aperture imaging runtimes for $L=12$ windows each spanning $40^o$ with $10^o$ overlap.}
\label{table:1}
\end{center}
\end{table}

Figure \ref{fig:contrast} shows a histogram of the $\log_{10}$ absolute value of various composite image estimates. The separation between the left and right ``humps'' of the histogram of the modulus values in each reconstruction displayed in Figure \ref{fig:contrast} further confirms that Algorithm \ref{alg:bcd} significantly increases contrast in the image while dampening background speckle. By contrast, the figure also shows that neither the NUFFT nor $\ell_1$ reconstructions sparsifies the image to the extent done by Algorithm 1.  In both cases, the left hump, which inherently represents pixels with no target reflectivity, is at least many orders of magnitude closer to zero. Since the goal of TV regularization is to sparsify its TV transform, it is not included in Figure \ref{fig:contrast}. 

Finally, Table \ref{table:1} shows another advantage -- the speed of reconstruction versus other sparsity encouraging methods, especially when the image size grows.  
Computations were done on a Macbook Pro with a 2.6 GHz 6-Core Intel Core i7 processor and 16GB memory. Implementation is achieved by using the NUFFT from \cite{fessler2003nonuniform}, as well as comparison code from \cite{sanders-imaging,sanders2017composite}.

%As mentioned in Section \ref{sec:background}, similar to the sampling method of \cite{CGspeckle} estimates of the speckle parameter $\boldsymbol{\alpha}$ and the noise variance $\beta$ result from Algorithm \ref{alg:sbl}. Indeed they are the components that make up the posterior variance estimate for $\mathbf{f}^{composite}$. However, since we derived MAP estimates for these quantities, uncertainty quantification is not available for them in this case.

%\begin{figure}[h]
%\centering
%\includegraphics[width=.49\textwidth]{images/mu_g}
%\caption{Comparison image using the sampling-based method of \cite{CGspeckle}, which is limited to direct full-azimuth imaging ($L=1$).}
%\label{fig:sampling}
%\end{figure}

%Finally, in Figure \ref{fig:sampling} we display the same example reconstructed using the sampling-based method of \cite{CGspeckle}, which as mentioned due to storage and memory requirements is limited to direct full-azimuth imaging method as opposed to composite (i.e. $L=1$). In addition, this method took $526$s to recover using $N=512^2$. However, this incorrect assumption also appears to adversely affect the results of the sampling method introduced in \cite{CGspeckle}, and in particular the magnitude of the true signal is not fully recovered as shown in Figure \ref{fig:sampling}.

\subsubsection{Total variation regularization testing}

Finally, as mentioned earlier, it is possible to incorporate a non-identity regularizer into Algorithm 1 as long as the kernels of the regularization matrix and the forward Fourier operator have trivial intersection. Due to the now sparse (as opposed to diagonal) covariance inversion problem that must take place at every step, this is a much more computationally expensive problem. Further efforts will be devoted to accelerating this task if these early tests show promising results. Therefore, as a proof of concept, we limit the problem size to $512\times512$ pixels, where we focus on a subregion of the parking lot from the same GOTCHA data set, \cite{casteel2007challenge}. To account for the smaller image size, we use $L=60$ sub-aperture windows each spanning $8^o$ with $2^o$ of overlap.

\begin{figure}[h]
\centering
\includegraphics[width=.3\textwidth]{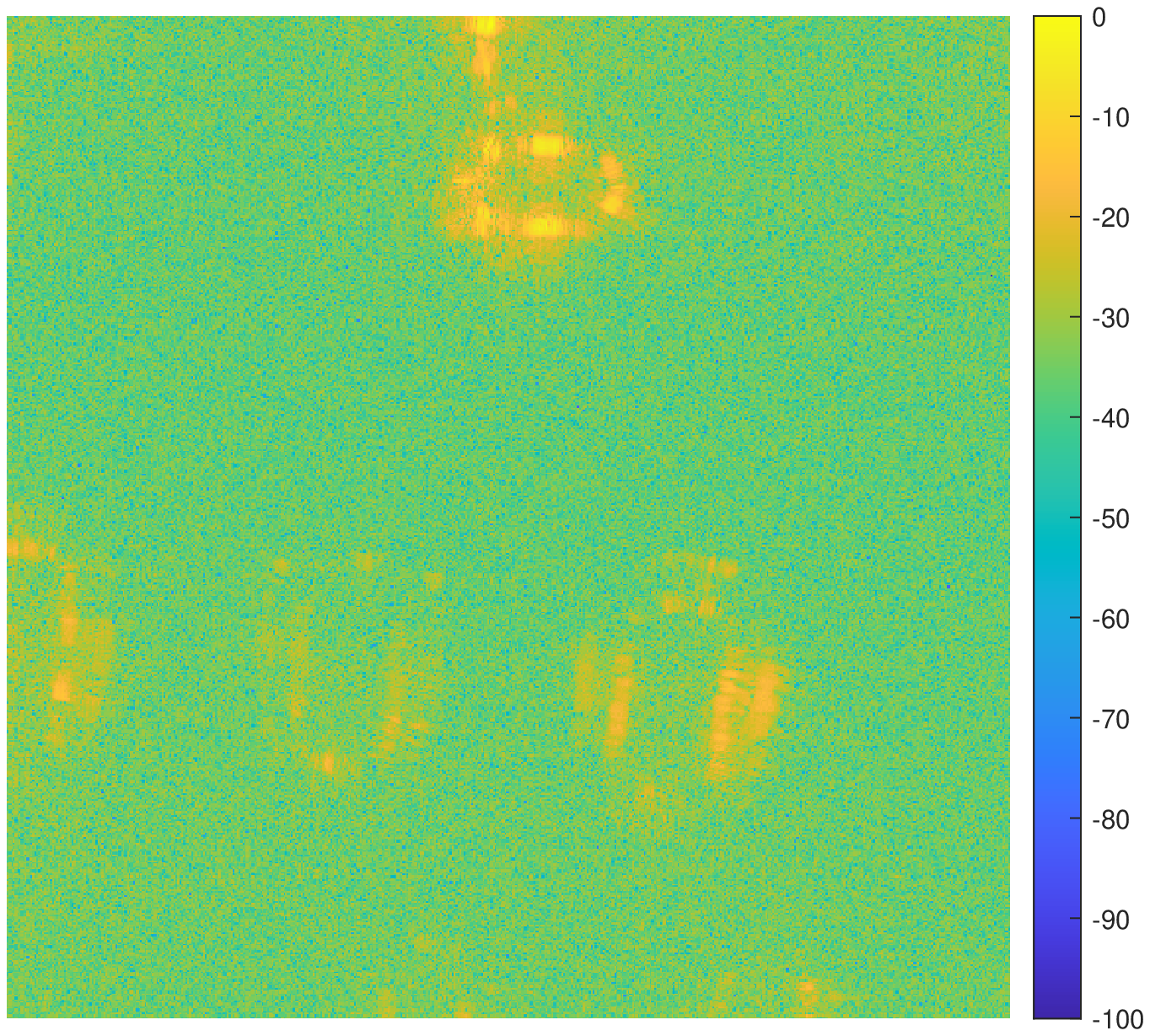}\includegraphics[width=.3\textwidth]{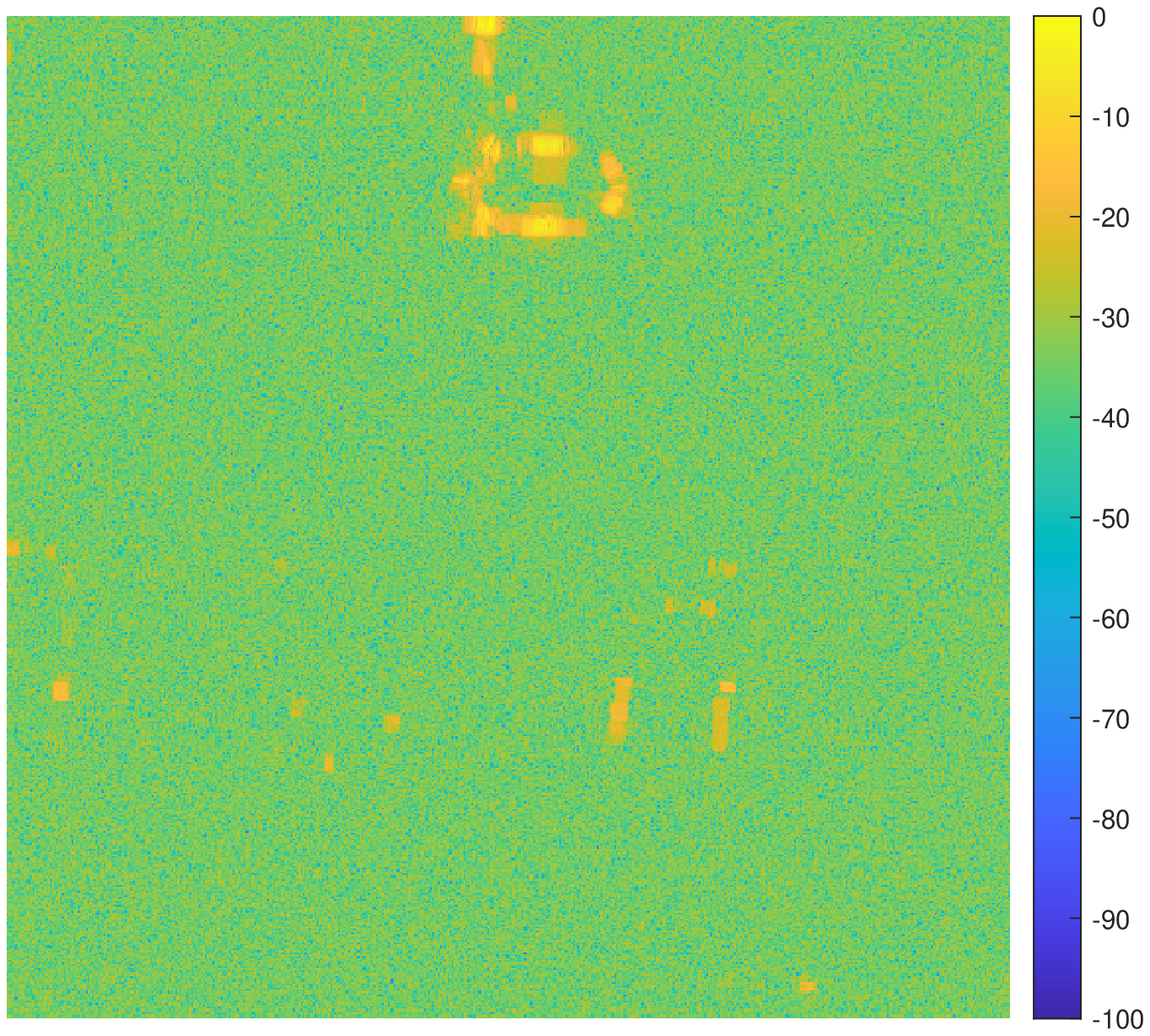}
\includegraphics[width=.3\textwidth]{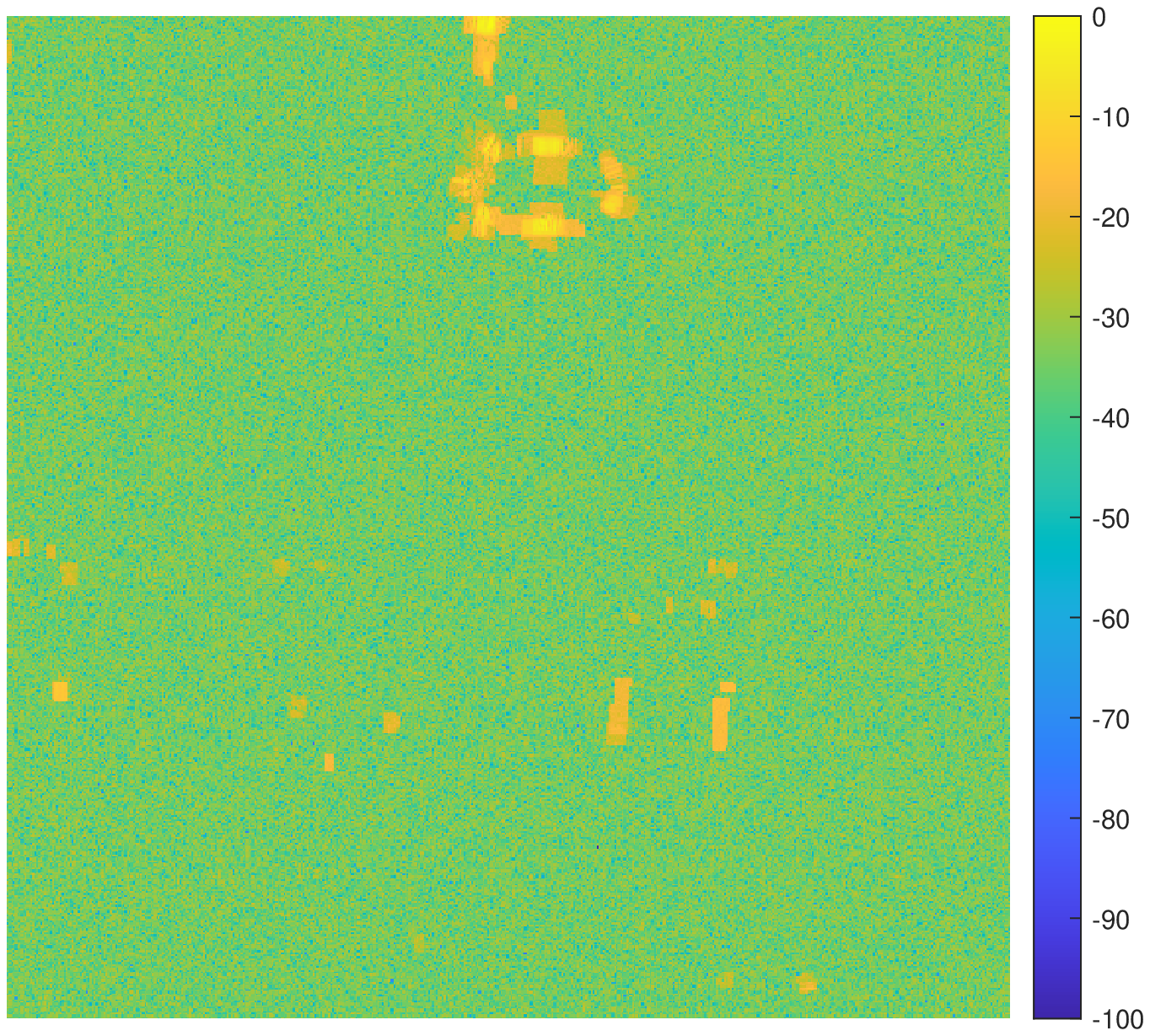}
\caption{Sub-aperture composite mean images formed with Algorithm \ref{alg:bcd} using the two-dimensional anisotropic TV operator for $\mathbf{T}$ and (left) $\epsilon=0.5$, (center) $\epsilon=0.1$, and (right) $\epsilon=0.01$ from the GOTCHA data set, \cite{casteel2007challenge}.}
\label{fig:GOTCHA_TVmean}
\end{figure}

\begin{figure}[h]
\centering
\includegraphics[width=.3\textwidth]{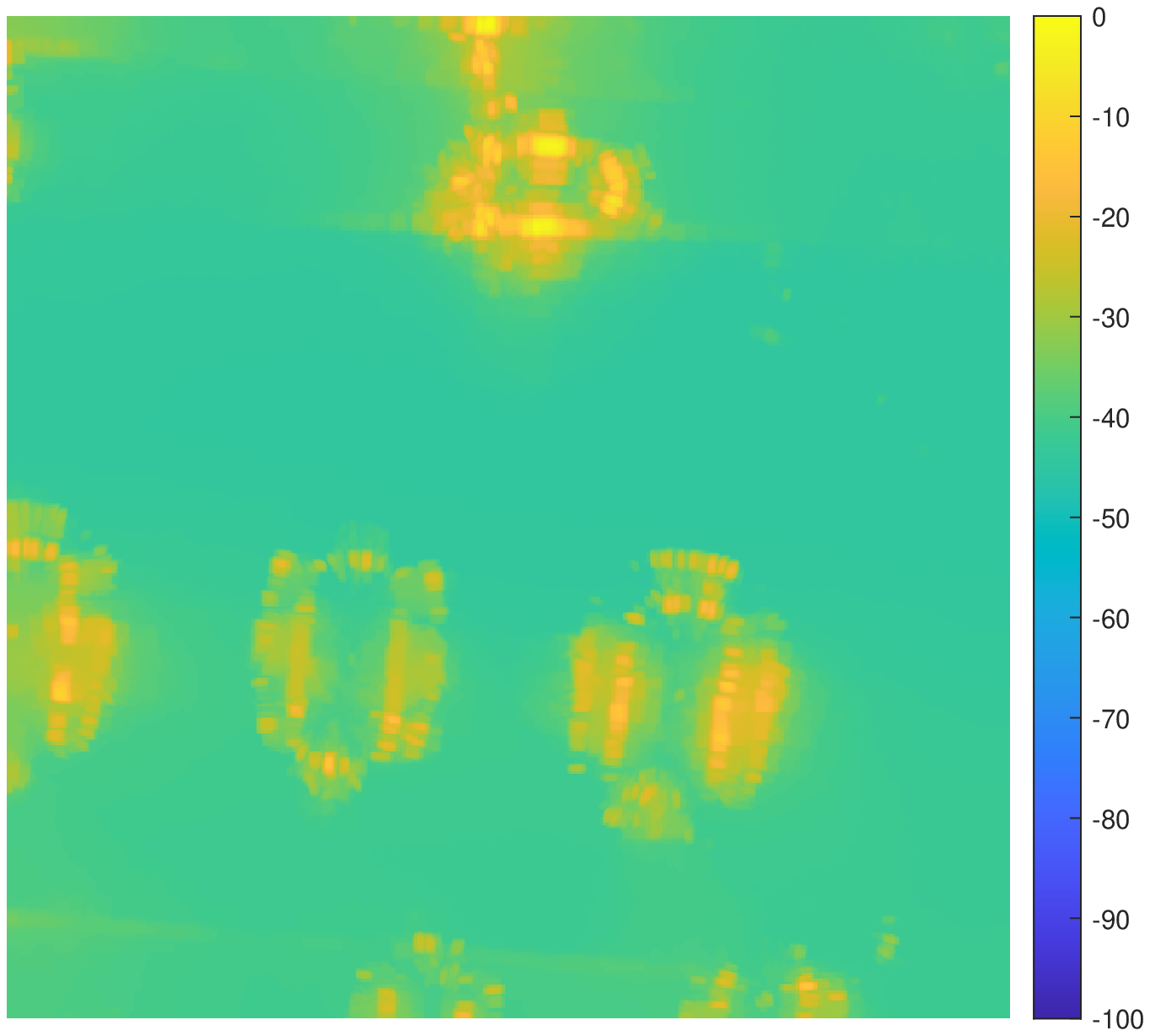}\includegraphics[width=.3\textwidth]{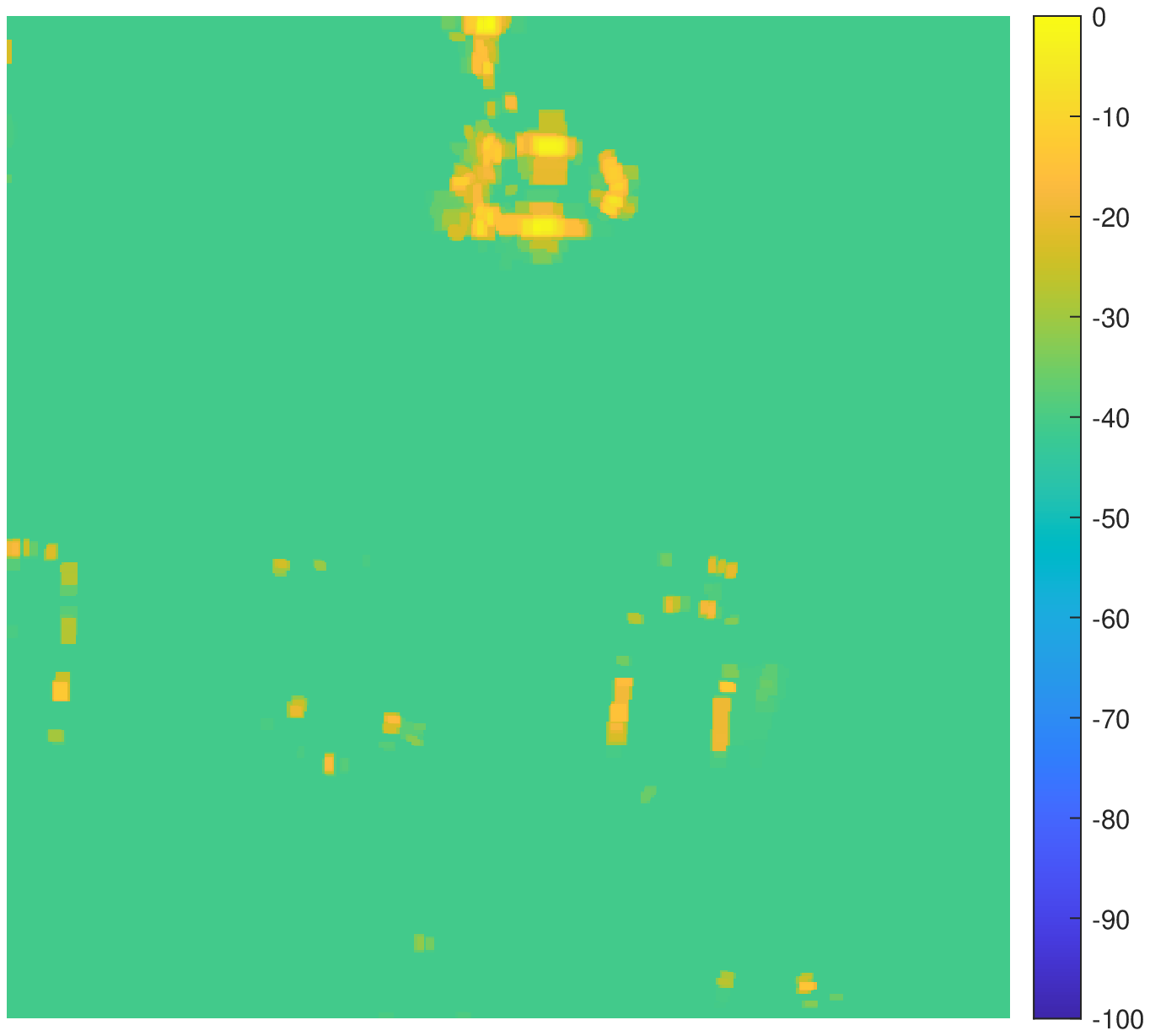}
\includegraphics[width=.3\textwidth]{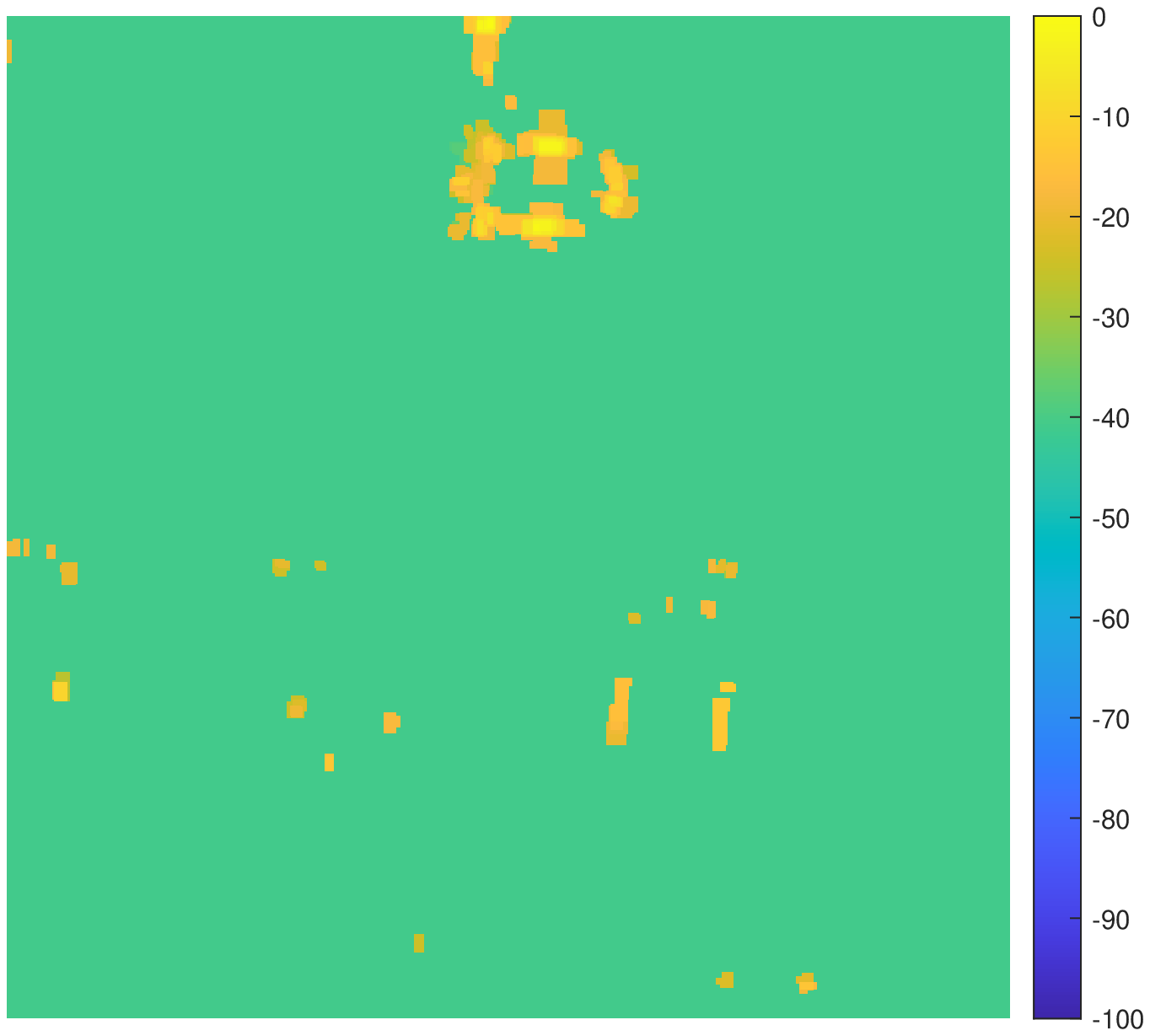}
\caption{Sub-aperture composite max images formed with Algorithm \ref{alg:bcd} using the two-dimensional anisotropic TV operator for $\mathbf{T}$ and (left) $\epsilon=0.5$, (center) $\epsilon=0.1$, and (right) $\epsilon=0.01$ from the GOTCHA data set, \cite{casteel2007challenge}.}
\label{fig:GOTCHA_TVmax}
\end{figure}

Figures \ref{fig:GOTCHA_TVmean} and \ref{fig:GOTCHA_TVmax} show mean and max composite estimates of the subregion using two-dimensional anisotropic TV regularization. We see some of the typical smoothing behavior associated with TV, although further testing is required to determine the speckled appearance of the mean composite image compared with the max composite image as well as the oversmoothing of important image features.

Uncertainty quantification information for the TV case also requires further investigation. In particular, further testing is also needed in assigning physical meaning to the ``speckle parameter'' $\boldsymbol{\alpha}$ here, as it does not represent speckle in this context. In addition, similar to the computation of Algorithm 1 itself with a non-identity regularizer, the computation of the composite covariance is still expensive even at this limited size so a standard deviation image is not available.

\section{Conclusion}\label{sec:conclusion}
The sub-aperture imaging approach separately analyzes small aperture windows of data to form image estimates that can combine into a single composite reconstruction. This  can help to mitigate the effects of conflicting information resulting from the anisotropic nature of the scatterers in the scene. In this paper, a hierarchical Bayesian prior and corresponding SBL-tyle estimation procedure is used to develop a method that models sub-aperture posterior densities as opposed to producing point estimates. This extra uncertainty quantification information provides a more robust result for practitioners to be confident in, with access to estimates for the parameters governing speckle and noise, as well as regarding the standard deviation of image pixels. A new coherent composite image combination method based on summing independent Gaussians is also presented. Furthermore, the method allows for a variety of admissible non-identity regularization operators such as TV. In terms of the estimates generated, we demonstrate with a real-world numerical example that compared to existing methods, the proposed algorithm reduces speckle, improves contrast, and is orders of magnitude closer in speed to the fast NUFFT method which allows large images to be processed.
Future work will focus on increasing efficiency as well as further exploring using non-identity regularizers in this framework. We also plan to incorporate autofocusing techniques into our method.

\bibliography{refs}
\bibliographystyle{acm}

\end{document}